\documentclass[12pt]{article}
\pdfoutput=1

\usepackage[margin=2cm]{geometry}
\usepackage[utf8]{inputenc}
\usepackage{graphicx}

\DeclareGraphicsExtensions{.jpg,.pdf,.mps,.png} 

\usepackage{fancyhdr}
\usepackage{color} 
\usepackage{apacite}

\newcommand{\bfx}{\mathbf{x}}

\begin{document} 

\begin{flushleft}
{\Large
\textbf{Solving constraint-satisfaction problems with distributed neocortical-like neuronal networks}
}
\footnote{This article has been accepted for publication in Neural Computation (MIT Press) on 01/12/18. This is the final peer-reviewed version that will appear in a future issue of Neural Computation.}

\vspace{1cm}

Ueli Rutishauser$^{1,2\ast}$, Jean-Jacques Slotine$^{3}$, Rodney J. Douglas$^{4}$
\\
\vspace{0.5cm}
$^1$ Computation and Neural Systems, Division of Biology and Biological Engineering, California Institute of Technology, Pasadena, CA, USA
\\
$^2$ Cedars-Sinai Medical Center, Departments of Neurosurgery, Neurology and Biomedical Sciences, Los Angeles, CA, USA
\\
$^3$ Nonlinear Systems Laboratory, Department of Mechanical Engineering and Department of Brain and Cognitive Sciences, Massachusetts Institute of Technology, Cambridge, MA, USA
\\
$^4$ Institute of Neuroinformatics, University and ETH Zurich, Zurich, Switzerland
\\
\vspace{1cm}
$\ast$ E-mail: Corresponding author urut@caltech.edu
\end{flushleft}

\section*{Abstract}
Finding actions that satisfy the constraints imposed by both external inputs and internal representations is central to decision making. We demonstrate that some important classes of constraint satisfaction problems (CSPs) can be solved by networks composed of homogeneous cooperative-competitive modules that have connectivity similar to motifs observed in the superficial layers of neocortex. The winner-take-all modules are sparsely coupled by programming neurons that embed the constraints onto the otherwise homogeneous modular computational substrate. We show rules that embed any instance of the CSPs planar four-color graph coloring, maximum independent set, and Sudoku on this substrate, and provide mathematical proofs that guarantee these graph coloring problems will convergence to a solution. The network is composed of non-saturating linear threshold neurons. Their lack of right saturation allows the overall network to explore the problem space driven through the unstable dynamics generated by recurrent excitation. The direction of exploration is steered by the constraint neurons. While many problems can be solved using only linear inhibitory constraints, network performance on hard problems benefits significantly when these  negative constraints are implemented by non-linear multiplicative inhibition. Overall, our results demonstrate the importance of instability rather than stability in network computation, and also offer insight into the computational role of dual inhibitory mechanisms in neural circuits.

\section{Introduction}	
The ability to integrate data from many sources, and to make good decisions for action is essential for perception and cognition, as well as for industrial problems such as scheduling and routing. The process of integration and decision is often cast as a constraint satisfaction problem (CSP). In technological systems CSPs are solved by algorithms that implement strategies such as back-tracking, constraint propagation, or linear optimization. In contrast to those algorithmic methods, we explore here the principles whereby neural networks are able to solve some important classes of CSP directly through their asynchronous parallel distributed dynamics. 

Our view of an algorithm here is computational one: The defined sequence of discrete operations to be carried out on data by a computer to achieve a desired end. By contrast, we view our neural networks as modeled approximations to physical neuronal networks such as those of the cerebral cortex, in which processing is not under sequential algorithmic control. Our primary interest in this paper is to understand how these natural networks can process CSP-like problems. 

We explore the behavior of these networks in the context of some reference classes of CSP that are well understood from an algorithmic point of view: 4-coloring of planar graphs (GC4P); minimum independent set (MIS); and the game Sudoku (SUD). For each of these classes the network must decide a suitable color assignment for each node of the graph given a total number of available colors (which is given a priori).  We choose these graph-coloring problems because their topologies lend themselves to implementation in networks, and because their constraints can be expressed as simple equal / not-equal relations.  Moreover, four-coloring on planar graphs is an interesting problem because it is computationally hard, and because its solution can be applied to practical tasks in decision making and cognition \cite{Afek2011, dayan2008simple, koechlin2007}. On the other hand, Sudoku is interesting because it involves many constraints on a dense non-planar graph \cite{Ercsey2012chaos,Rosenhouse2011taking}. Also, there are hard input constraints on the values of some SUD nodes, which makes the solution of SUD significantly more difficult than simple graph coloring, in which any valid coloring is acceptable. 

We show that our neuronal networks can solve arbitrary instances in these three problem classes. Because many problems in decision making can be reduced to one of these classes \cite{dayan2008simple} showing that our networks can solve them, implies that they can in principle solve all other problems that are reducible to these reference classes. These problems can of course be solved also by algorithmical methods \cite{Kar72,Robertson1996,AppHakKoc77,kumar1992algorithms}. However, our important contribution here is to explain the principles of dynamics that allow a network of distributed interacting neurons to achieve the same effect without relying on the centralized sequential control inherent in these well known algorithms.    
    
Our networks are composed of loosely coupled cooperative-competitive, "winner-take-all" (WTA) modules \cite{Hahnloser00,Maass00,Hahnloser99,Douglas2007_recurrent,YuilleGeiger03,Ermentrout92,wersing2001competitive,Abbott94}. Each WTA behaves as a special kind of variable that initially is able to entertain simultaneously many candidate values, but eventually selects a single best value, according to constraints imposed by the processing evolving at related variables. This collective selection behavior is driven by the signal gain developed through the recurrent excitatory (positive feedback) connections between the neurons participating in the WTA circuits, as described below.

WTA circuits are appealing network modules because their pattern of connectivity resembles the dominant recurrent excitatory and general inhibitory feedback connection motif measured both physiologically \cite{douglas_intracellular_1991} and anatomically \cite{binzegger_quantitative_2004} in the superficial layers of the mammalian neocortex.  There is also substantial and growing evidence for circuit gain and its modulation in the circuits of the neocortex, a finding that is consistent with the recurrent connectivity required by WTAs \cite{douglas_canonical_1989,douglas_intracellular_1991,ferster_orientation_1996,lien_tuned_2013,li_linear_2013, carandini_normalization_2012, Kaminski2017}.

Previous studies of WTA networks have focused largely on questions of their stability and convergence (For example \cite{ben-yishai_theory_1995,AbbottDayan01, Hahnloser00,RutishauserDouglas2011}).  However, more recently we have described the crucial computational implications of the unstable expansion dynamics inherent in WTA circuits \cite{rutishauser_computation_2015}. It is these instabilities that drive the selection processing and therefore the computational process that the network performs. We have explored this computational instability in networks of linear thresholded neurons (LTN) \cite{Koch1998,douglas_role_1999,Lecun2015deep} because they have unbounded positive output. Consequently, networks of LTNs with recurrent excitation must rely on feedback inhibition rather than output saturation (eg \cite{Hopfield82, MillerZucker1999, RosenfeldZucker1976}) to achieve stability. This also means that networks of LTNs may change their mode of operation from unstable to stable \cite{rutishauser_computation_2015}.

The principles we develop in this paper depend on concepts we have described previously, in which we apply contraction theory \cite{Slotine03} to understand the dynamics of collections of coupled WTAs \cite{RutishauserDouglas2011, rutishauser_computation_2015}. And so, for convenience, we first summarize briefly the relevant points of that work. First, note that contraction theory is applicable to dynamics with discontinuous derivatives such as those introduced by the LTN activation function that we utilize. This is because in such switched systems the dynamics remain a continuous function of the state (see results for details). Under suitable parameter regimes, LTN networks can enter unstable subspaces of their dynamics. The expansion within an unstable subspace of active neurons is steered and constrained by the negative divergence of their dynamics. This ensures that the current expanding sub-network will soon switch to a different sub-network, and so to a different subspace of dynamics \cite{rutishauser_computation_2015}. The new space might be stable or unstable. If unstable, the network will again switch, and so on, until a stable solution space is found. We refer to the unstable spaces as 'forbidden' spaces because the network will quickly exit from them. The exit is guaranteed because the dynamics in forbidden spaces are such that the eigenvectors associated with the dominant eigenvalue of the system Jacobian are always mixed. This means that the activity of at least one neuron is falling toward threshold, and will soon pass below it, so changing the space of network dynamics (see \cite{rutishauser_computation_2015} for a proof.) Stable spaces, on the other hand, are said to be be 'permitted'. In these spaces the eigenvalues are all negative and so the network will converge toward a steady state in that space. 

Critically, negative divergence ensures that the dynamics of the space entered next has a lower dimensionality than the previous space, regardless of whether it is stable or unstable. Thus the sequence of transitions through the sub-spaces causes the network to compute progressively better feasible solutions. A further crucial feature of this process is that the direction of expansion is determined automatically by the eigenvectors of the Jacobian of the currently active neurons in the network. Thus, the direction of expansion may change according to the particular set of neurons active in a given forbidden subspace. In this sense the network is able to actively, asynchronously, and systematically search through the forbidden spaces until a suitable solution subspace is encountered. It is this process that constitutes network computation  \cite{rutishauser_computation_2015}.

Now we extend these concepts and show how they can be utilized to construct networks that solve certain classes of constraint satisfaction problems. We show using new mathematical proofs and simulations how such problems can be embedded systematically in networks of WTA modules coupled by negative (inhibitory) and positive (excitatory) constraints. Our overall approach is to ensure that all dynamical spaces in which a constraint is violated are 'forbidden'. This is achieved by adding additional neurons that enforce the necessary constraints. Importantly, our new analytical proofs guarantee that these networks will find a complete and correct solution (provided that such a solution does exist for the problem instance). 

We also find that the form of inhibitory mechanism used to implement negative constraints affects the performance of the network. Two different types of inhibition can be used to implement negative constraints: linear subtractive and non-linear multiplicative inhibition. While some problem classes could be solved using only standard subtractive inhibition between modules, we found that the solution of more difficult problems is greatly facilitated by using multiplicative non-linear inhibition instead. Recent experimental observations have implicated these two kinds of inhibition in different modes of cortical computation \cite{Jiang2013organization,Pi2013cortical}, and the work presented here offers a theoretical foundation for the computational roles of these two types of inhibition also in the neuronal circuits of the neocortex.

\section{Network Architecture and Results}

\subsection{CSP Organization}
A CSP consists of a set of variables (concepts, facts) $X_i$ that are assigned discrete or continuous values and a set of constraints $C_i$ between these variables. The constraints may be unary (restricted to one variable), binary (a relation between two variables), or higher order. Each constraint $C_i$ establishes the allowable combinations between values (or relationships between values) of the variables $X$. A state (or configuration) of the problem is an assignment of values to some or all of the variables $X$. An assignment (or solution) may be complete or partial.

The CSPs are instantiated as neuronal networks by embedding the specific problem in a field of identical WTA modules. These modules have a standard connection pattern of recurrent excitation and inhibition that supports the WTA functionality. The CSP is embedded in the network by coupling the WTA modules via neurons that implement the negative (inhibitory) and positive (excitatory) constraints. As we will show below, we find that the performance of the CSP network is affected by the form of negative constraint inhibition onto the WTAs; either linear or non-linear. We begin by describing the 'standard' WTA (WTA$^S$) and related CSP networks that make use of linear inhibitory negative constraints; and thereafter describe the extended WTA (WTA$^E$) networks that make use of non-linear inhibitory negative constraints.      

\subsection{Standard Winner-Take-All circuit}
\begin{figure}
\centering
\includegraphics[angle=0,width=18cm]{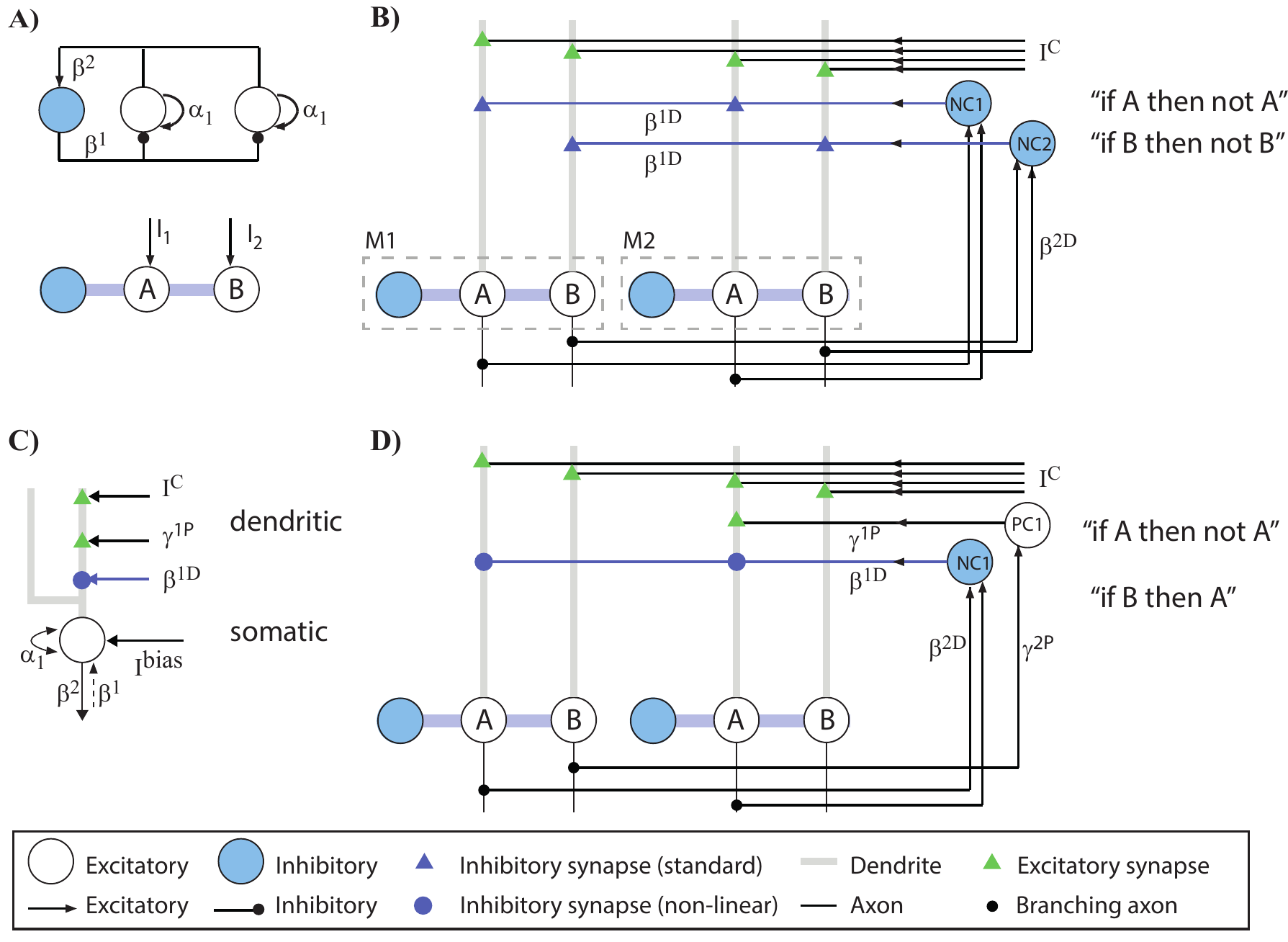} 
\caption{Connectivity and network architecture. 
(A) Single WTA comprising two excitatory neurons that encode possible winners, A and B. Top: all connections. Bottom: simplified notation.
(B) Two modules M1 and M2 that implement WTA shown in (A) are connected with one another by two additional inhibitory cells $NC_{1,2}$. These cells enforce the negative constraint that the two WTAs cannot reach the same winner (solution). The constraint inhibition is linear. To maintain analogy with neurons, the inhibition is shown applied to a dendrite. However, in this point model neuron, it could as well be applied directly to the soma. See Fig \ref{fig:simpleSim} for a simulation of this constraint problem.
(C) The non-linear inhibitory synapse (blue circle) provides on-path inhibition, which can suppress only dendritic but not somatic excitatory inputs.
(D) Example circuit with both (non-linear) negative, and positive (NC, PC) constraint cells implemented with non-linear inhibitory synapses.
}
\label{fig:circuit1}
\end{figure}

Each standard WTA (WTA$^S$) consists of N point neurons, N-1 of which are excitatory and the remaining one (N) is inhibitory (Fig \ref{fig:circuit1}A). In the examples below the WTA should express only a single active unit, and thus the excitatory neurons $x_{i \ne N}$ receive only self-feedback of excitation $\alpha_i$. Each neuron may receive also external input $I_i$. Normally distributed noise with mean $\mu$ and standard deviation $sd$ is added to all these external inputs: $I_i = I_i + \mathcal{N}(\mu,sd))$. 

The single inhibitory neuron $x_N$ sums the $\beta^2$ weighted input from each of the excitatory neurons, and returns a common $\beta^1$ inhibitory signal to all of its excitatory neurons. The dynamics of this single WTA are

\begin{equation} \label{eq:wta-excite}
\tau \dot{x}_i + G x_i = f(\alpha x_i - \beta^1 x_{N} - T_i + I_i + \mathcal{N}(\mu,sd)) \\
\end{equation}
\begin{equation} \label{eq:wta-inhib}
\tau \dot{x}_{N} + G x_{N}  = f( \beta^2\sum_{j=1}^{N-1} x_j - T_N)
\end{equation}
\noindent where $f(x)$ is a non-saturating rectification non-linearity. Here, we take $f(x)=max(x,0)$, making our neurons linear threshold neurons (LTNs). $T_i \geq 0$ is the activation threshold.

\subsection{Constraint satisfaction models implemented on WTAs}
The nodes of the graph represent the possible states of the problem at that location, while the edges of the graph represent the constraints acting between nodes. Each node is a WTA module, and the patterns of activation of its excitatory neurons represent the allowable states of that node. Since only one winner is permitted, these WTAs represent as many solution states as they have excitatory neurons, ie $N-1$ states. In the CS problems considered here, all nodes implement the same set of states. For example, in the case of GC4P, every node of the problem graph is a WTA with 4 excitatory neurons that represent the four possible color states of that node. 

The constraints between WTAs are implemented by additional neurons that add additional excitatory or inhibitory interactions between the relevant states (corresponding to particular neurons) of the interacting nodes. In graph-coloring problems, the constraints are negative: The selection of a particular color neuron at one node, suppresses the selection of the corresponding color neuron at a neighboring node. However, other problems may require also positive constraints. For example, if WTA A is in state 1, then WTA B should also be in state 1. The Maximal Independent Set (MIS) problem considered below requires such positive constraints. In this section, we will first describe the dynamics and connectivity of negative-and positive contraint cells whereas their specific wiring to implement a particular CSP is described later separately for each CSP class considered.

Negative constraints (NC), or competition, between the states of different WTAs is introduced through inhibitory feedback by negative constraint cells (NCC) $d$. In our implementation, each $d$ provides inhibitory output onto the same set of excitatory cells from which it receives its input (Fig \ref{fig:circuit1}B). In contrast to the inhibitory cells that enforce competition between states within an WTA, NC cells enforce their competition between (some) specific state neurons across multiple participating WTAs. 

Positive constraints (PC), or hints, are implemented by excitatory positive constraint cells (PCC) $p$, each of which receives input from a specific excitatory cell of one WTA and provides excitation onto specific excitatory neurons(s) on other WTAs.

The dynamics of negative constraint cells $d$ are:
\begin{equation}
\tau \dot{d}_{i} + G d_{i}  = f( \beta^{2_D}\sum_{j=1}^{M_i} x_j - T^D)
\label{eq:SimpleInhibConstraint}
\end{equation}
\noindent where $M_i$ are the number of units $x_j$ that provide input to $d_i$. Note that the $x_j$ are members of different WTAs. Similarly, the dynamics of a positive constraint cell $p$ are:
\begin{equation}
\tau \dot{p}_{i} + G p_{i}  = f( \gamma^{2_P}\sum_{j=1}^{N_i} x_j - T^P)
\label{eq:SimpleExcitConstraint}
\end{equation}
\noindent where $N_i$ are the number of units $x_j$ that provide input to $p_i$. 

\noindent The total constraint current, composed of negative and positive components $I^{NC}_i$, and $I^{PC}_i$ is:

\begin{equation}
I^{cons}_i = I^{NC}_i + I^{PC}_i = -\sum_{k=1}^{D_i} \beta^{1_D}  d_k + \sum_{k=1}^{P_i} \gamma^{1_P} p_k 
\label{eq:TotalSimpleConstraints}
\end{equation}
\noindent where $D_i$ and $P_i$ are the total number of negative and positive constraint cells that provide input to cell $i$, respectively.
$\beta^{1_D}>0$ is the strength of the inhibitory synapse made by negative constraint cell $d_k$ onto cell $i$ and $\gamma^{1_P}>0$ is the strength of the excitatory synapse made by positive hint cell $p_k$ onto cell $i$.

Thus, the dynamics of excitatory units in constraint networks composed of standard WTA $x_i$ are:
\begin{equation}
\tau \dot{x}_i + G x_i = f_s( \alpha x_i - \beta^1 x_{N} - T_i  + I_i + I^{bias}_i + I^{cons}_i )
\label{eq:SimpleConstraints}
\end{equation}

where $I^{bias}_i$ are forward inputs that bias the WTA towards solution $x_i$. The initial conditions for all units are $x_i(0)=0, d_i(0)=0, p_i(0)=0$ throughout this work.

\subsection{Dynamics of an example negative constraint network}
\begin{figure}
\centering
\includegraphics[angle=0,width=18cm]{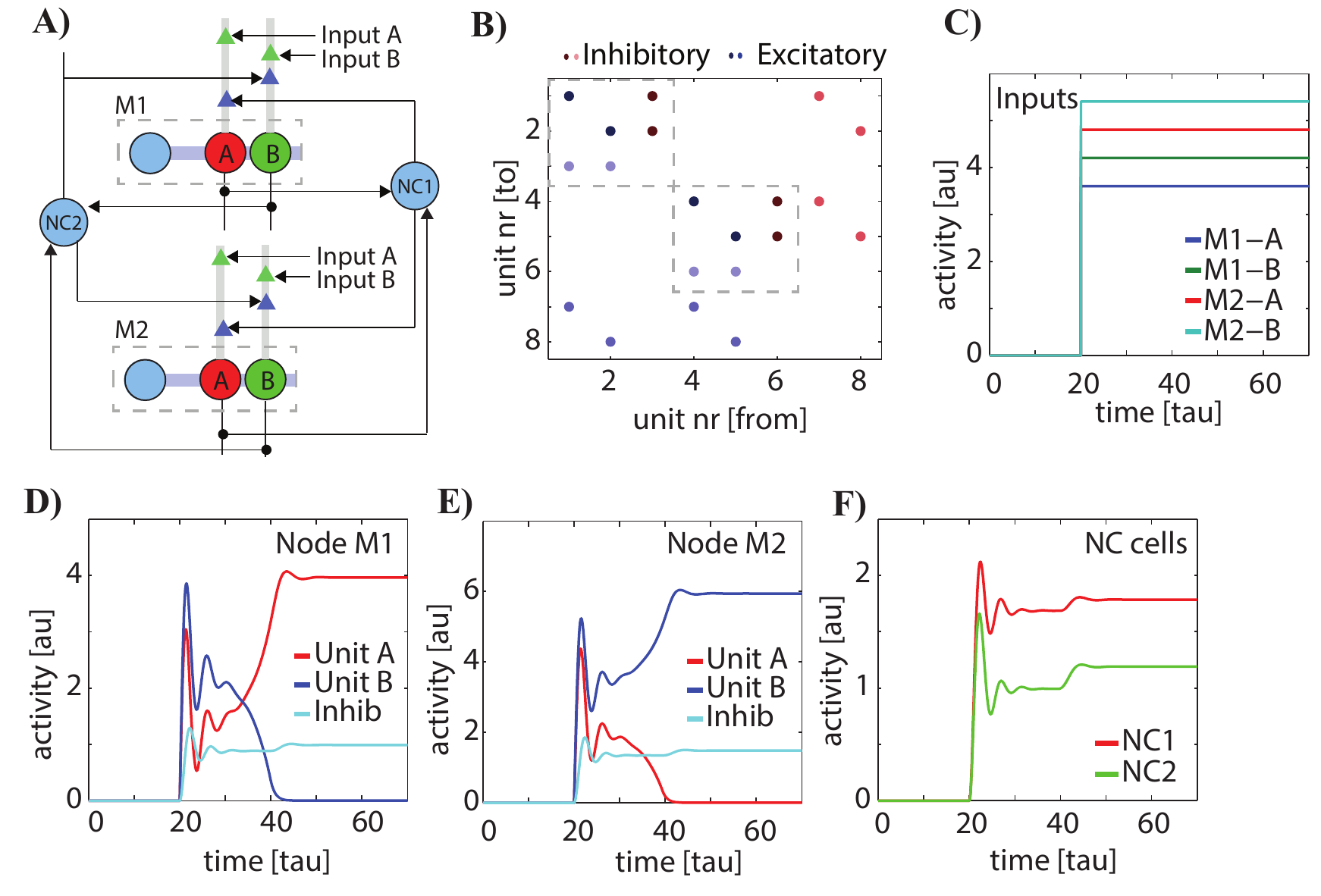} 
\caption{Enforcing constraints between WTAs using negative constraint (NC) cells. Simple example using the two-Node circuit of Fig \ref{fig:circuit1}B.
(A) Circuit diagram of the network, with two nodes M1 and M2 with two winners A and B each. The two negative constraint cells NC enforce the not-same constraint.
(B) Weight matrix of the full network. Gray boxes mark nodes M1 and M2. Connections outside of the boxes correspond to the NCs.
(C) The inputs to the network. 
(D,E) Activity on nodes M1 and M2. (F) Activity of the two negative constraint cells NC1 and NC2. 
}
\label{fig:simpleSim}
\end{figure}

Before we consider how to choose the parameters of this network and how to analyze its stability and convergence, consider the example network shown in Fig \ref{fig:circuit1}B. This network implements two negative constraints (NC1 and NC2) between two WTA modules, M1 and M2. Each has only two possible solutions: A and B. The NC cells enforce the "not same" constraint that both WTAs should not be in the same state. The dynamics of this circuit (Fig \ref{fig:simpleSim}) converge to a steady-state in which the solutions of the two WTAs depend on each other in addition to local constraints (the inputs). For example, in Fig \ref{fig:simpleSim} unit $B$ on both WTAs receive the largest inputs (cyan,red) and so, independently, the winner on either WTA would be $B$. However, because of the constraint dependency only one WTA can express $B$ (node M2), whereas the winner on the other node (M1) is $A$ despite receiving the lower amplitude input. 

\subsection{Stability and convergence} 
Our arguments rely on the concept of the effective Jacobian, which expresses the dynamics of the currently active subset of neurons \cite{rutishauser_computation_2015}. Consider the  network in the form of $\dot \mathbf{x} = z(\mathbf{x},t)$. Here, $z(\mathbf{x},t)=f(\mathbf{W}\mathbf{x}-\mathbf{G}\mathbf{x})$, where $\mathbf{W}$ is a matrix of weights that describes all the connections between neurons, $f(x)$ is the non-linearity, and $\mathbf{G} = {\rm diag}(G_1,\dots,G_n)$ is a diagonal matrix containing the dissipative leak terms for each neuron. The effective Jacobian for this system is
\begin{equation}
\mathbf{J}_{eff} = \frac{\partial \mathbf{z}} {\partial \bfx} = \Sigma \mathbf{W}  - \mathbf{G}
\label{eq:Jeff}
\end{equation}
\noindent where $\Sigma = {\rm diag}(\sigma_1,\dots,\sigma_n)$ is a diagonal matrix of derivatives of the activation function, evaluated at the current state $x_i$ of each neuron. In our case, all neurons are LTNs, resulting in derivatives equal to either $0$ or $1$ depending on whether the state of a neuron is above or below threshold. This is because the activation function $f(x)$ has slope $1$ whenever it is above threshold. Therefore, $\Sigma$ remains the same as long as no neuron crosses its threshold. Consequently, the effective connectivity of the network changes whenever a neuron crosses its threshold. 'Effective connectivity' indicates that connections that arise from inactive (below threshold) neurons cannot influence their target neurons because they do not provide output. Therefore, the rows of the Jacobian corresponding to these silent neurons are zeroed out. However, their columns are not zeroed, and so these silent neurons still receive and process input.

We use $\mathbf{J}_{eff}$ as a mathematical tool to assess and describe network computation. We have used $\mathbf{J}_{eff}$ previously to show that a single WTA circuit has both permitted (contracting) and forbidden (expanding) sets of active neurons, and that it is the existence of the forbidden sets that provides computational power \cite{rutishauser_computation_2015}. This is because forbidden sets are highly unstable and this drives the network to enter a different set from the one it is currently expressing. While unstable, the negative feedback ensures that the dynamics during forbidden states steers network activity towards a suitable solution. A forbidden set must satisfy two conditions: its divergence must be negative, and its effective Jacobian $\mathbf{J}_{eff}$ must be positive definite \cite{rutishauser_computation_2015}. In our WTA networks, these two conditions are enforced by shared inhibition and excitatory self-recurrence, respectively.  Together, they guarantee that an individual WTA will exit forbidden sets exponentially fast. 

The arguments summarized above and the new results derived below rest on contraction theory \cite{Slotine03}, a powerful analytical tool that allows us to systematically reason systematically about the stability and instability of non-linear networks such as the LTN networks that we use. Contraction theory is applicable to non-linear networks, such as switched networks, provided that the dynamics remain a continuous function of the state, and that the contraction metric remains the same \cite{Lohmiller2000}. It is thus applicable to networks composed of units such as LTNs that have activation functions whose derivatives are discontinuous. To see why this is the case, consider  the following network: $\dot{x} + x = max(x,0)$. Note that $\dot{x}$, which describes the dynamics of the network, is a continuous function of $x$ despite having discontinuous derivatives with respect to $x$. Furthermore, the metric of our networks remains the same throughout their processing \cite{rutishauser_computation_2015}. Thus, both conditions for the application of contraction theory are satisfied.

Now we present new proofs that together provide important insights into the operation of this network. First, we prove that adding NC cells creates new forbidden sets. The NCs are thereby able to influence the dynamics of network computation. Second, we prove that adding PC cells creates permitted sets. Third, we prove that networks of WTAs connected by NC/PC cells remain stable despite the presence of forbidden sets. Together, these new results generalize our previous findings from individual WTAs \cite{rutishauser_computation_2015} to networks of WTAs coupled by NC and PCs that implement specific constraints. Finally, we provide rules that allow all instances of three classes of CSPs to be implemented in networks of WTAs by installing suitable NC and PC coupling connections.

\subsubsection{Proof 1: Adding negative constraint cells creates forbidden sets}

Consider a set of WTAs, with the dynamics of each described by a Jacobian $\mathbf{J}_{i}$:
\begin{equation}
\mathbf{J}_{i} = \left[ 
\begin{array}{ccc}
l_1 \alpha-G & 0        & -\ l_1 \beta^1           \\
0        & l_2 \alpha-G & -\ l_2\beta^1           \\
l_3 \beta^2  & l_3 \beta^2  & -G                 \\
\end{array} 
\right]
\label{eq:Jmain}
\end{equation}

Now, consider a system composed of two copies of the above circuits $\mathbf{J}_{1,2}$:
\begin{equation}
\mathbf{J}_{3} = \left[ 
\begin{array}{cc}
\mathbf{J}_1 & \mathbf{0} \\
\mathbf{0}   & \mathbf{J}_2  \\
\end{array} 
\right]
\label{eq:combine1}
\end{equation}

$div(\mathbf{J}_3)<0$ if $div(\mathbf{J}_{1,2})<0$. Thus, combinations of circuits with negative divergence will always have negative divergence.  Next, we add an additional inhibitory neuron ("NC cell") that enforces additional competition between the two sub-circuits $\mathbf{J}_1$ and $\mathbf{J}_2$. This new constraint will create a new forbidden subspace, which must be expanding. This system is described by $\mathbf{J}_4$.

\begin{equation}
\mathbf{J}_4 = \left[ 
\begin{array}{cc}
\mathbf{J}_3 & -k \mathbf{D}^T \\
\mathbf{D}   & -G  \\
\end{array} 
\right]
\label{eq:combine3}
\end{equation}

with $k=\frac{\beta^1}{\beta^2}$. For example, setting $\mathbf{D} = \left[ \begin{array}{cccccc} \beta^2 & 0 & 0 & \beta^2 & 0  & 0 \end{array} \right]$ would connect the new inhibitory unit such that simultaneously activating the first unit in both sub-circuits $\mathbf{J}_1$ and $\mathbf{J}_2$ is forbidden. 

The individual WTA $\mathbf{J}_{1,2}$ is expanding if $\mathbf{V}_{1,2} \mathbf{J}_{1,2} \mathbf{V}_{1,2}^T>0$ with
\begin{equation}
\mathbf{V}_{1,2} = \left[ 
\begin{array}{ccc}
1 & -1 & 0
\end{array} 
\right]
\label{eq:V12}
\end{equation}
\noindent This V term ensures that both excitatory units on a WTA cannot be simultanously active.

The combined system $\mathbf{J}_{4}$ is expanding if $\mathbf{V}_{4} \mathbf{J}_{4} \mathbf{V}_{4}^T>0$ with
\begin{equation}
\mathbf{V}_4 = \left[ 
\begin{array}{ccc}
\mathbf{V} & \mathbf{0} & \mathbf{0} \\
\mathbf{0} & \mathbf{V} & \mathbf{0} \\
\mathbf{V}_j & -\mathbf{V}_j & \mathbf{0}
\end{array} 
\right]
\label{eq:combine4}
\end{equation}
\noindent where $\mathbf{V}_j= \left[ \begin{array}{ccc} 1 & 0 & 0 \end{array} \right]$. The term $[\mathbf{V}_j -\mathbf{V}_j]$ ensures that the first excitatory
neurons on both WTAs cannot be simultaneously active, which is the constraint that the NC cell described by $\mathbf{D}$ embeds. Multiplying out, the result becomes
\begin{equation}
\mathbf{F}_{4} = \mathbf{V}_{4} \mathbf{J}_{4} \mathbf{V}_{4}^T = \left[ 
\begin{array}{ccc}
\mathbf{V} \mathbf{J}_1 \mathbf{V}^T   & 0                                       & \mathbf{V}_j \mathbf{J}_1 \mathbf{V}^T \\
0                                      & \mathbf{V} \mathbf{J}_1 \mathbf{V}^T    & -\mathbf{V}_j \mathbf{J}_2 \mathbf{V}^T \\
\mathbf{V}_j \mathbf{J}_1 \mathbf{V}^T & -\mathbf{V}_j \mathbf{J}_2 \mathbf{V}^T & \mathbf{V}_j \mathbf{J}_1 \mathbf{V}_j^T + \mathbf{V}_j \mathbf{J}_2 \mathbf{V}_j^T
\end{array} 
\right]
\label{eq:combine5}
\end{equation}
\noindent which can further be decomposed into
\begin{equation}
\mathbf{F}_{4} = \left[ 
\begin{array}{cc}
\mathbf{Q}_1 & \mathbf{B}^T \\
\mathbf{B}   & \mathbf{Q}_2 
\end{array} 
\right]
\label{eq:combine6}
\end{equation}
\noindent A sufficient condition for above to be positive definite is, $\mathbf{Q}_2>\mathbf{B}^T \mathbf{Q}_1^{-1} \mathbf{B}$ \cite{Horn85} (Page 472).
Note that this condition requires $\mathbf{Q}_{1,2}$ to be symmetric. After substituting all variables, the result is

\begin{equation}
\mathbf{F}_{4} = \left[ 
\begin{array}{ccc}
-2+2\alpha & 0 & -1+\alpha \\
0          & -2+2\alpha & 1-\alpha \\
-1+\alpha  & 1-\alpha & -2+2\alpha
\end{array} 
\right]
\label{eq:combine5subs}
\end{equation}
\noindent Thus, $\mathbf{Q}_2=2(\alpha-1)$, $\mathbf{B}=[-1+\alpha, 1-\alpha]$, and $\mathbf{Q}_1=2(\alpha-1) \mathbf{I}$.
Accordingly, $\mathbf{F}_{4}$ is positive definite if $2(\alpha-1) > \alpha-1$. This condition is satisfied if $\alpha>1$.

In conclusion, the additional feedback loop introduced by adding the negative constraint cell creates a forbidden subspace, which is both negatively divergent as well as expanding. This method can be applied recursively to add arbitrary numbers inhibitory feedback loops to a collection of WTA circuits.

Positive constraints do not create additional forbidden sets. Instead, a positive constraint of the kind "if WTA1 is in state 1, WTA2 should be in state 2" reinforces two sets that are already permitted. Thus, all that is required with regard to positive constraints, is a proof demonstrating that permitted sets remain permitted (see below).

\subsubsection{Proof 2: Stability analysis}
\label{proof:stability}
In this section, we demonstrate that the addition of positive and negative constraints does not disturb the overall stability of the system.
A key feature of contracting systems is that contraction is preserved through a variety of combinations of subsystems \cite{Slotine03}. Importantly, this property includes the two kinds of combination that we consider here: the introduction of negative and positive constraint cells. In the following, we simply outline the grounds for this inclusion. The detailed proof of preservation of contraction under combination of subsystems is presented in \cite{Slotine03}. 

We begin by showing that the addition of negative constraint NC cells does not disturb the stability of the system. Coupling two WTAs $\mathbf{J}_1$ and $\mathbf{J}_2$ with 
NC cells such that the two WTAs can not have the same winners results in the system Jacobian $\mathbf{J}_3$ (see Equation \ref{eq:combine1}).
\begin{equation}
\tau \mathbf{J} = \left[ 
\begin{array}{cc}
\mathbf{J}_{3} & -k \mathbf{B}^T \\
\mathbf{A}     & -\mathbf{G} \\ 
\end{array} 
\right]
\label{eq:Jfullneg}
\end{equation}
\noindent where the NC cells have no dynamics apart from the load term $G$ on the diagonal. The connectivity of the NC cells is
$\mathbf{A}=\left[\begin{array}{cccccc} \beta^{2D} & 0 & 0 & \beta^{2D} & 0 & 0 \end{array}\right]$. 
With $k=\frac{\beta^{1D}}{\beta^{2D}}$, $\mathbf{B}=\mathbf{A}$. Feedback combinations of this form are guaranteed to be contracting as shown in \cite{Slotine03} (section 3.4).

A similar argument holds for positive constraint cells: Adding a positive constraint between two identical WTAs results in a new system with Jacobian
\begin{equation}
\tau \mathbf{J} = \left[ 
\begin{array}{cc}
\mathbf{J}_{3} & \mathbf{B}^T \\
\mathbf{A}     & -\mathbf{G} \\ 
\end{array} 
\right]
\label{eq:Jfullpos}
\end{equation}
\noindent where the positive constraint cell has no dynamics apart from the load term $g$ on the diagonal. For the example of a single
positive constraint cell that enforces that if WTA 1 is in state 1, WTA 2 should be in state 2, its connectivity is: 
$\mathbf{A}=\left[\begin{array}{cccccc}\gamma^{2P} & 0 & 0 & 0 & 0 & 0 \\\end{array}\right]$ and 
$\mathbf{B}=\left[\begin{array}{cccccc} 0             & 0 & 0 & 0 & \gamma^{1P} & 0 \\\end{array}\right]$.
Unlike the previous negative constraint case, there is no simple relationship between $\mathbf{A}$ and $\mathbf{B}$. However, taking the symmetric part $\mathbf{J}_S = \frac{1}{2}(\mathbf{J}+\mathbf{J}^T)$ results in a system in which $\mathbf{A}_S=\mathbf{B}_S=\left[\begin{array}{cccccc}\frac{\gamma^{1P}+\gamma^{2P}}{2} & 0 & 0 & 0 & \frac{\gamma^{1P}+\gamma^{2P}}{2} & 0 \\\end{array}\right]$. Feedback combinations of this form are guaranteed to be contracting \cite{Slotine03} (section 3.4) if
\begin{equation}
\sigma^2(\mathbf{A}_S) < \lambda(\mathbf{J}_3)\lambda(\mathbf{-G})
\end{equation}
\noindent where $\sigma$ is the largest singular value and $\lambda$ the largest eigenvalue. Both $\lambda(\mathbf{J}_3)$ and $\lambda(\mathbf{-G})$ are negative by definition, since the systems are both individually contracting. 

The contraction rate of a WTA is $\frac{\alpha}{2}-1$ (see \cite{rutishauser_competition_2012} section 2.3.2). Assuming $G=1$, this reduces to
\begin{equation}
(\gamma^{P1}+\gamma^{P2})^2<2-\alpha
\label{eq:gammaConstraintContraction0}
\end{equation}
\noindent Therefore, as long as the weights of the positive constraint cells fulfill condition Equation \ref{eq:gammaConstraintContraction0}, the system will remain contracting.

\subsection{Choice of parameters}
\label{section:paramChoice}
The stability and computational power of a WTA depends on the following parameters: excitatory local recurrence $\alpha$, inhibitory recurrence $\beta^1$ and $\beta^2$ (note that $\beta^2$ is a superscript index and not a power), and inhibitory $\beta^{{1,2}_D}$ and excitatory ($\gamma^{P1}$,$\gamma^{P2}$) recurrence between WTAs that implements the constraints. Provided these parameters are set to values within a permitted range, this network will allow only one winner to emerge, and that solution depends on the pattern of its input $I$ \cite{RutishauserDouglas2009}. These constraints are: 

\begin{equation}
1<\alpha<2 \sqrt{\beta_1 \beta_2 }
\label{eq:standard1}
\end{equation}
\begin{equation}
\frac{1}{4}<\beta_1\beta_2<1
\label{eq:standard2}
\end{equation}
\begin{equation}
\beta_1 \beta_2 < (1 - \frac{1}{\alpha})(\beta_1^2 + \frac{\alpha^2}{2})
\label{eq:standard3}
\end{equation}
\begin{equation}
(\gamma^{P1}+\gamma^{P2})^2<2-\alpha
\label{eq:gammaConstraintContraction}
\end{equation}

where Equations \ref{eq:standard1}-\ref{eq:standard3} are derived in \cite{RutishauserDouglas2011} and Equation \ref{eq:gammaConstraintContraction} is derived above. Note that the constraints on inhibitory feedback loops established by inhibitory neurons apply to both the local inhibitory neuron of each WTA, as well as the additional inhibitory neurons that establish inhibitory feedback between WTAs (referred to as $\beta_{1,2}$ and $\beta^{{1,2}_D}$, respectively).

For all simulations, we chose parameters within the permitted ranges given by Equations \ref{eq:standard1}-\ref{eq:gammaConstraintContraction}. Within those restrictions, the parameters were chosen to optimize performance for each problem class (i.e. GC4P, MIS, and SUD) and network type ($WTA^{s}$ and $WTA^{e}$).  Note that the parameters used were identical for all instances of a particular problem class and network type (i.e. GC4P solved with the $WTA^{s}$ architecture) and not optimized for a particular problem instance (which are randomly generated).

\subsection{Planar Graph Coloring using only negative constraints (GC4P)}
\begin{figure}
\centering
\includegraphics[angle=0,width=18cm]{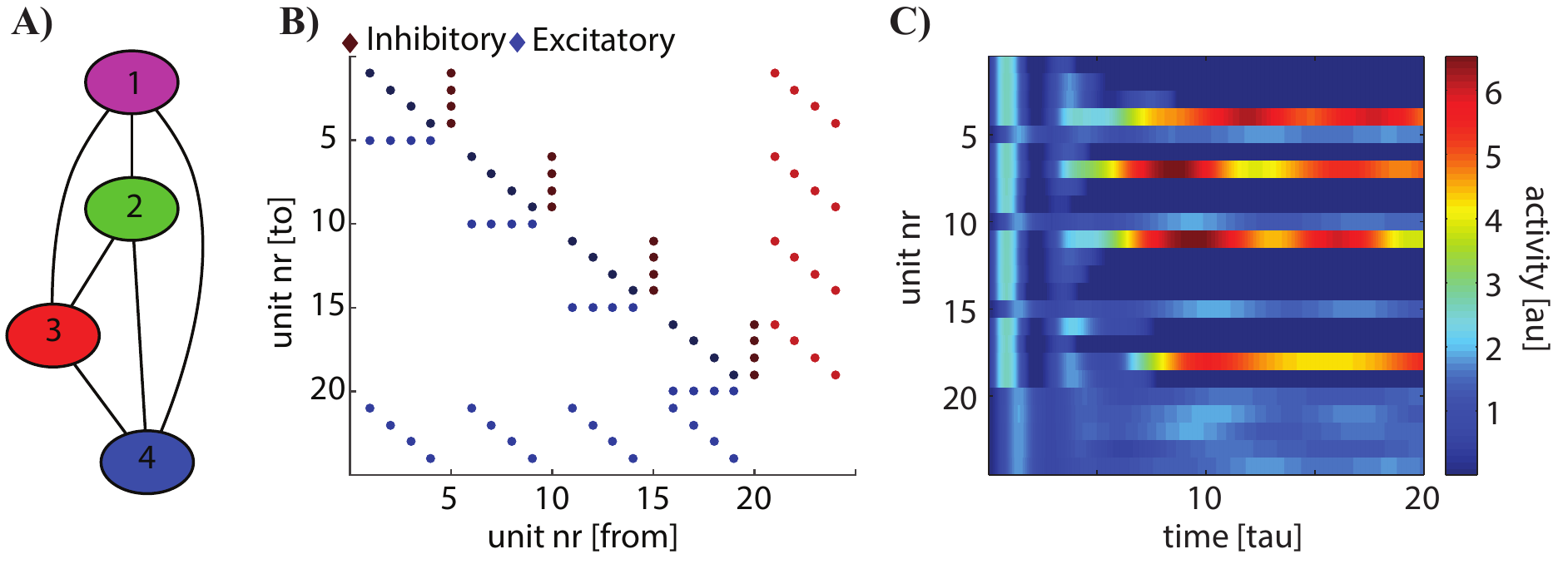} 
\caption{Solving graph coloring problems with networks of WTAs.
(A) Example 4-node graph, with one possible color solution computed by a WTA network indicated (colors).
(B) Weight matrix of the 4 module WTA network, implementing the graph shown in (A). Neurons 1 through 20 are configured as 4 separate WTAs, each with four excitatory neurons that encode the 4 possible colors of each node, and one global (to that WTA) inhibitory neuron. Neurons 21-24 impose the inhibitory constraint that no edge may have nodes of the same color. 
(C) Dynamics of the network leading to solution in (A). Note the relatively small modulations of constraint neuron activity required to achieve this solution.}
\label{fig:graphcolor-simple}
\end{figure}

We next applied this architecture to the problem of graph coloring. Here, each node must at any time express only one of a fixed number of different colors. The selected color represents the node's current state. The coloring constraint is that nodes that share an edge are forbidden from having the same color. Finding an assignment of colors to all the graph nodes that respects this constraint for all their edges solves the graph coloring problem (for example, Fig \ref{fig:graphcolor-simple}A). The smaller the number of permitted colors, the harder the problem.

\begin{figure}
\centering
\includegraphics[angle=0,width=18cm]{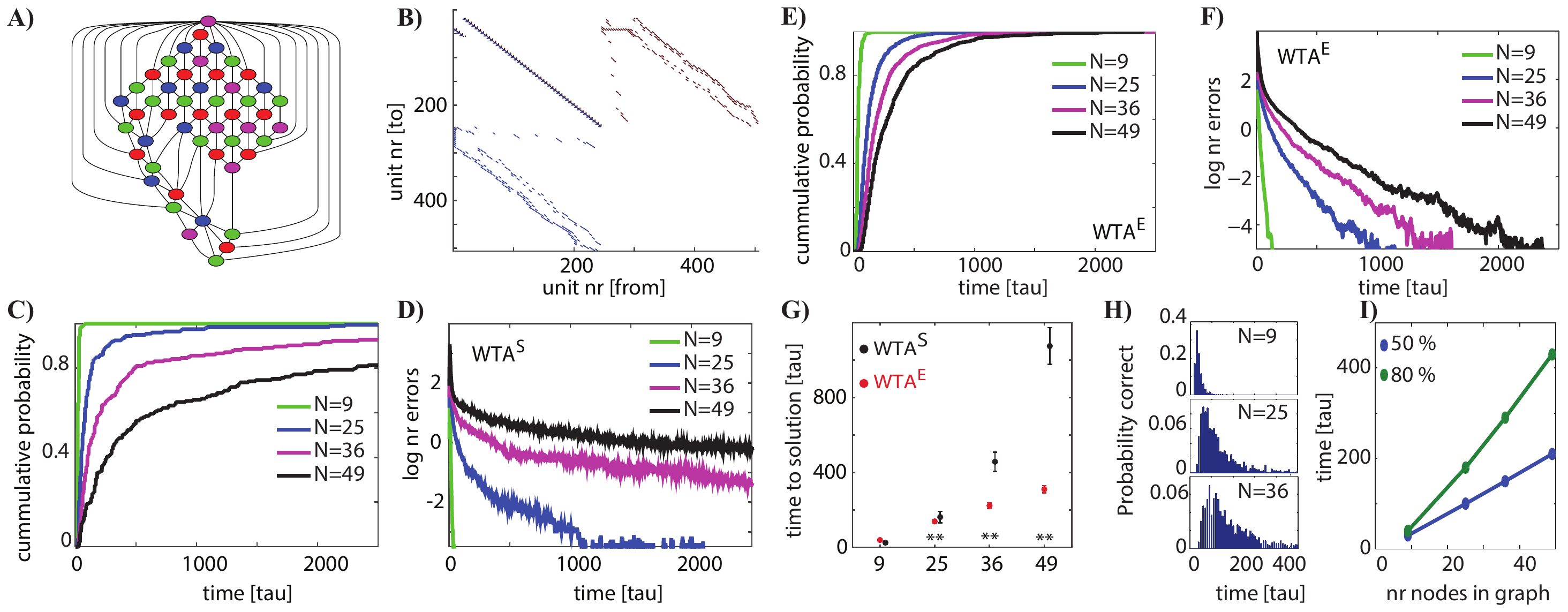} 
\caption{Performance on solving the planar graph four coloring problem (GC4P). 
(A) Example of a GC4P solution.
(B) Weight matrix of a network with 505 units (245 for WTAs and 260 NCs).
(C,D) Performance of $WTA^{s}$. (C) Cumulative probability of network convergence as a function of processing time and network size. 
(D) Average number of errors (graph edge constraints violated) as a function of time.
(E,F) Performance for $WTA^{e}$. Same notation as (C,D).
(G) Average time $\pm s.e.$ to find a solution as a function of network size and architecture. Time to solution for large problems was significantly shorter for $WTA^{e}$ network by comparison with $WTA^{s}$ (**, p<0.01, kstest). 
(H) Distribution of times to solution, as a function of network size.
(I) Scaling of time at which solved 50\% and 80\% of all networks converged as a function of network size. 
$WTA^{s}$ parameters: $\alpha=1.5, \beta^1=3, \beta^2=0.3, \beta^{1_D}=1.5, \beta^{2_D}=0.15$, iid noise of $\mu=1.5, \sigma=0.15$. 
$WTA^{e}$ parameters: same, except $\alpha=1.2, \beta^{1_D}=3, \beta^{2_D}=0.3, s=0.15, o=0$. 
See section \ref{section:paramChoice} for how the network parameters were chosen. Results are for N=1000 simulations for each network size. A new random planar graph with 80\% density was generated for each simulation. 
} 
\label{fig:graphcolor-all}
\end{figure}

More specifically, we chose to investigate the problem of coloring planar graphs with 4 colors with an arbitrary number of nodes and undirected edges (Figs \ref{fig:graphcolor-all}). A planar graph is one that can be embedded in the plane, which means that it can be drawn in such a way that no edges cross one another. Here, we restrict ourselves to planar graphs, because they are guaranteed to be colorable with 4 colors.

The topology of Graph Coloring can naturally be framed as a topologically distributed constraint satisfaction problem, and so implemented by networks of WTA circuits in the manner that we have described above. The color state of each node is represented by a single WTA, and so the network implementation requires as many WTA modules as the graph has nodes. In the problems reported here, the smallest number of colors required to color the graph (its chromatic number) is constant (here, 4) and given. Thus, each WTA has as many state neurons (possible winners) as the chromatic number. Because the local competition between these states is unbiased, all WTAs have the same internal connection architecture. 

The edges of the graph are constraints of the kind "not same", and they are implemented using NC cells (see Methods; Fig \ref{fig:graphcolor-simple}). At most, the constraints across each edge will require as many different NC cells as there are node colors. However, a single NC cell can enforce its constraint across arbitrary numbers of neighbor WTAs. Thus, it is sufficient to add only one NC cell per color at a given node. This cell is then able to assert the same constraint across all edges connected to this node (see Methods).

We have shown previously for both symmetrical \cite{Hahnloser00} and asymmetrical networks \cite{rutishauser_computation_2015} that the fundamental operation of the WTA modules is an active selection process whereby the activities of some neurons are driven below threshold by those who are receiving support from either local or remote excitatory input. Partitions of active neurons that are inconsistent with stability are forbidden. Such a partition is left exponentially quickly because the unstably high gain generated by the neurons of forbidden partitions will drive recurrent inhibition sufficiently strongly to soon drive a neuron of the set beneath threshold, and so bring a new partition into being. This process continues until a consistent permitted partition is found. The previous work was concerned only with the relationship between inhibitory feedback that is driven by the excitatory members of the local WTA, and the existence of forbidden sets. We demonstrated there that it is the existence of these forbidden subspaces that provides computational power \cite{rutishauser_computation_2015}. Now, in these CSP networks the negative constraints provide an additional source of negative feedback routed via remote WTAs (see above for a proof). 

We tested the performance of the network in solving randomly generated planar graphs with up to 49 four color nodes (Fig. \ref{fig:graphcolor-all}). In these cases there were no constraints on the acceptable color of any specific node. Thus, any solution that satisfies all constraints is acceptable. Consequently there are many equally valid solutions for each graph (at minimum four, the number of colors).

The goodness of a solution was measured by two metrics: The average time the network took to converge, and the number of edges that were not satisfied (Number of Errors, in the following) as a function of time. We found that WTA networks can solve all graphs of up to 25 nodes correctly within 1500$\tau$. However, larger graphs are incompletely solved and take longer (Fig \ref{fig:graphcolor-all}C). The computational process evolves in such a way that the number of errors decreases exponentially fast (Fig \ref{fig:graphcolor-all}D). This means that networks of a given size solve the majority of random graphs quickly, with a minority taking much longer. 
This results in heavy-tailed distributions with respect to the elapsed time to solution (Fig \ref{fig:graphcolor-all}H). 

\subsection{Maximal Independent Set using both positive and negative constraints (MIS)}
\begin{figure}
\centering
\includegraphics[angle=0,width=13cm]{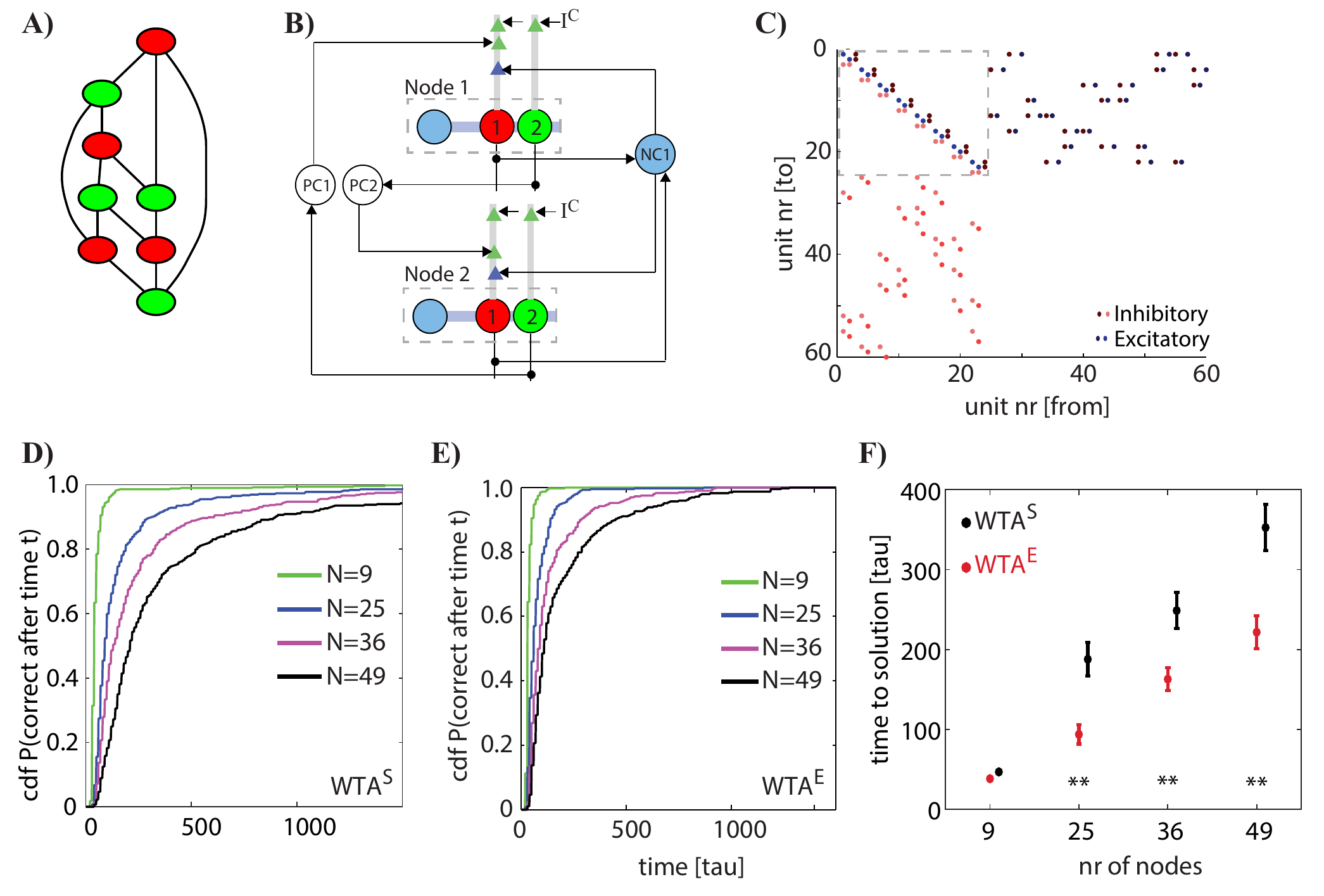} 
\caption{Performance on solving the Maximal Independent Set (MIS) problem.
(A) Example maximal independent set (red nodes) on an 8 node graph. Each red node is not connected to any other red nodes and each green node is connected to at least one red node.
(B) Connectivity for a simple two-node problem. Each node has two possible winners (red, green). NC1 enforces that not both notes can be red. PC1 and PC2
enforces that if a node is green, the other is red.
(C) Weight matrix for the 8-node graph illustrated in (A). The dashed box indicates the connection submatrix of the 8 WTAs (3 units each). Remaining entries indicate constraint units and their connections. 
(D-E) Performance of $WTA^{s}$ (D) $WTA^{e}$ (E) on random MIS problems of different size (number of nodes). Graphs were randomly generated planar graphs with 90\% density. $WTA^{e}$ converges more quickly than $WTA^{s}$ for all problem sizes.
(F) Performance comparison between $WTA^{s}$ (black) and $WTA^{e}$ (red). Except for the smallest problem (size N=9), $WTA^{e}$ converged significantly more quickly than $WTA^{s}$ (** is p<0.01, ks-test).
$WTA^{s}$ parameters: $\alpha=1.2, \beta^1=3, \beta^2=0.3, \beta^{1_D}=1.5, \beta^{2_D}=0.15, \gamma^{1_P}=0.8, \gamma^{2_P}_2=0.15$, iid noise of $\mu=1.5, \sigma=0.15$. 
$WTA^{e}$ parameters: Same, except $\gamma^{1_P}=1.5$. 
} 
\label{fig:MIS-all}
\end{figure}

The use of positive constraint cells is demonstrated by solving a second class of graph coloring problems: maximal independent sets (MIS)(Fig \ref{fig:MIS-all}). MIS problems are a second fundamental class of computational problems \cite{Afek2011} solvable with the type of network we present here. MIS is related to graph coloring, but requires different constraints. In this case, each node must take one of two possible colors A and B. If two nodes are connected (they are neighbors), then they cannot both be A. Further, any node that takes color B must be connected to at least one other node that has color A. This problem finds practical application in many distributed algorithms \cite{Lynch1996_distributed}, where it is used for automatic selection of local leaders. 

We translated MIS problems into networks that use both negative and positive constraints. The first constraint, that two neighbors can not both be color A, is implemented by one NC cell for each connected pair (see Fig \ref{fig:circuit1}D). The second constraint, that if a node has color B it encourages its neighbors to be of color A, is implemented by positive feedback through two hint cells for each pair of connected nodes (see Fig \ref{fig:circuit1}D). The positive feedback is active conditional on a node being of color B. Thus, if a node has color A, the positive constraint is inactive. We found that this WTA network solves MIS in a manner and speed similar to that described above for graph coloring (Fig \ref{fig:MIS-all}D): The networks solve most MIS problems of large size fairly quickly, however a small number of large problems remain unsolved even at long times. 

\subsection{Fully constrained graph coloring problems - Sudoku (SUD)}
The standard WTA networks are also able to solve non-planar graph coloring problems in which the color states of many nodes have a fixed assignment. These initial assignment constraints make graph coloring significantly more difficult than the case in which any valid solution is acceptable. A canonical example of this problem class is the popular game Sudoku. 

\begin{figure}
\centering
\includegraphics[angle=0,width=12cm]{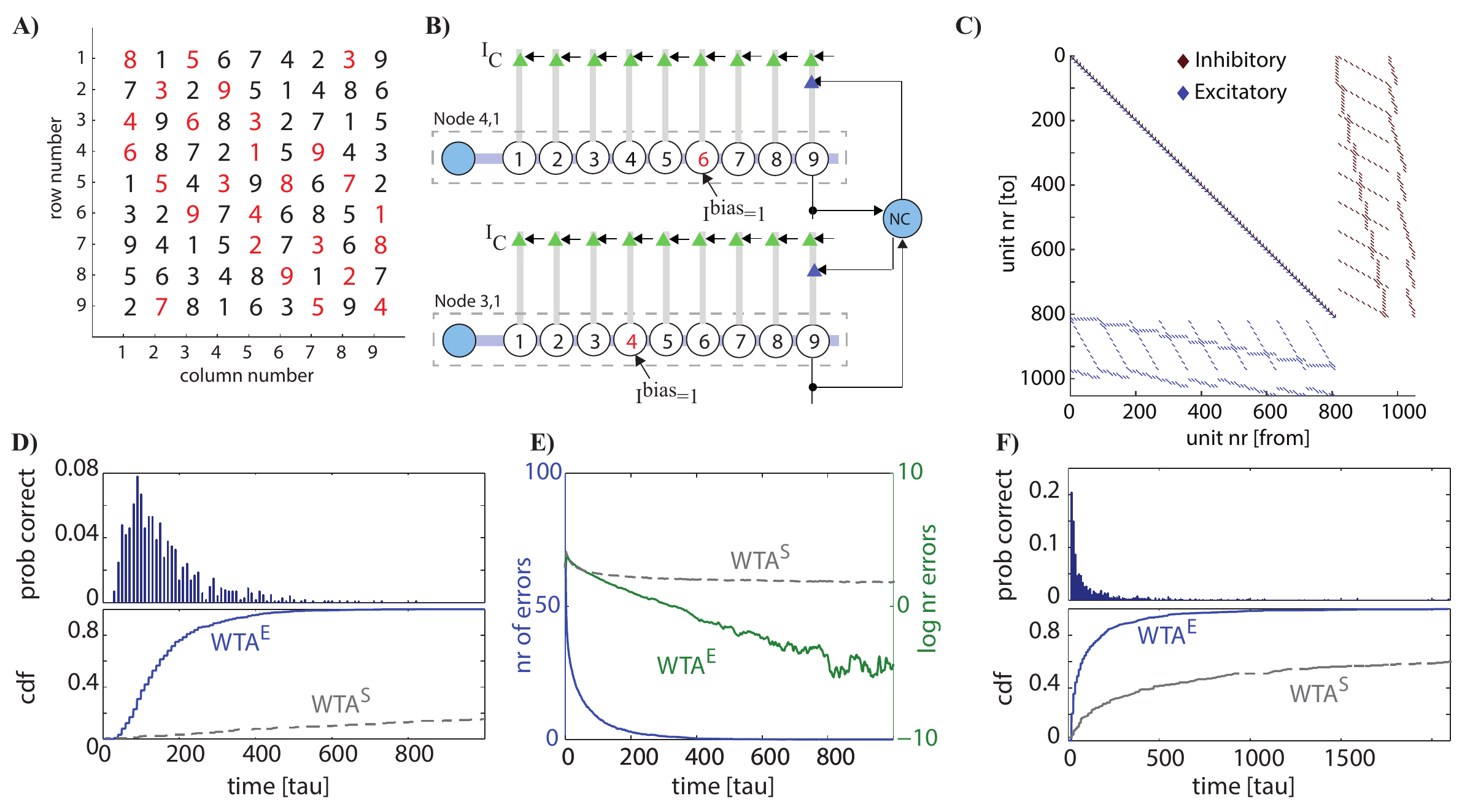} 
\caption{Sudoku, a graph coloring problem, solved by $WTA^{e}$.
(A) Example "hard" SUD, identical to the "hard" example used in \protect\cite{Habenschuss13}. Red values are given.
(B) Circuit implementation of SUD. Each node has 9 possible winners (colors). Row, column and box constrains are enforced through negative constraint (NC) cells). The pre-defined (red) winners are enforced through bias currents to the soma.
(C) Weight matrix of network that implements SUD. The network consists of 1052 units. Of those, 810 units (81 nodes, 9 excitatory and 1 inhibitory each) implement the nodes (WTAs) and 243 implement the constraints (9 row, 9 column and 9 box constraints, one for each color; i.e. 27*9).
(D) Performance of the $WTA^{e}$ (blue) and $WTA^{s}$ (gray) network on the sudoku shown in (A). 1000 runs of the same network with different initial conditions. $WTA^{e}$ required on average 161 $\tau$ to converge, with a maximal duration of ~800$\tau$. 
(E) Number of violated constraints (errors) decreases exponentially as a function of simulation time for the simulations shown in (D). 
(F) Same as in (D), but for simulations of 50 different sudoku problems of varying difficulty \protect\cite{ProjectEuler}. For $WTA^{e}$ and $WTA^{s}$, average convergence time was 142 $\tau$ and 1330$\tau$, respectively. $WTA^{e}$ parameters: $\alpha=1.1, \beta^1=3, \beta^2=0.3, \beta^{1_D}=3, \beta^{2_D}=0.3, s=4, o=4$. All contextual inputs $I^C=4$, with s.d. of $1$. $WTA^{s}$ parameters were identical except $\alpha=1.5, \beta^{D_1}=1.5, \beta^{D_2}=0.15$ (see Fig \ref{fig:graphcolor-all}). 
}
\label{fig:sudoku-results}
\end{figure}

The graph of Sudoku has 81 nodes arranged in a 9x9 lattice (Fig \ref{fig:sudoku-results}A,B). The lattice is composed of 3x3 boxes, each of which has 3x3 nodes. Each node can take one of nine colors (numbers). The constraints of the problem are that each color can appear only once in each row, in each column, and in each box. The neural network that implements Sudoku consists of 1052 units. Of those, 810 units (81 nodes, 9 excitatory and 1 inhibitory each) implement the nodes (WTAs) and 243 implement the constraints (9 row, 9 column and 9 box constraints, one for each color; i.e. 27*9). In addition, there are initial constraints in the form of specified inputs to a subset of the nodes which describe the specific problem to be solved (Fig \ref{fig:sudoku-results}A). 

For the sudoku network, each excitatory cell receives a constant noisy excitatory input from the network. This input stands for the broader network context in which the particular CS network is embedded. In addition, each excitatory cell receives several negative constraint inputs. There are no positive hint constraints. The forward inputs $I^{bias}_i$ enforce the fixed color assignments for some nodes, as required by the specific SUD problem to be solved. Unbiased neurons receive $I^{bias}_i=0$. 

We found that these standard $WTA^{s}$ networks are able to solve many Sudoku problems. But their performance is poor. The average converge time was $1330\tau$, and the network was only able to solve 60\% of all networks in the maximal time permitted. As in the case of graph coloring, the number of violated constraints (errors) decreases exponentially as a function of simulation time (Fig \ref{fig:sudoku-results}E,F).

\subsection{Extended WTA ($WTA^{e}$) networks}
We sought to improve the fraction of correct solutions and rate of reduction in errors by modifying the network configuration. We reasoned that the exponential form of the network convergence reflects its exploration of the combinatorial solution space; and that convergence would be more rapid if the network could be made more sensitive to its constraints. 

We noticed that in equation \ref{eq:SimpleConstraints} the mechanism of state selection within a WTA is similar to that of negative constraints; both operate via subtractive inhibition (i.e. the term $-\beta^{1_D} d_j$ is added). Therefore a negative constraint, rather than only discouraging the selection of a particular state, could prevent a state from ever being selected: If the subtractive inhibition through negative constraints cells is sufficiently strong, this cell will be driven far below the activation threshold and so become insensitive to positive inputs. We hypothesized that, if the effects of the constraints scaled multiplicatively rather than being applied subtractively as in equation \ref{eq:SimpleConstraints}, then the network would be able to separate the function of state selection at a node from the constraints that biased that selection, and that this change in architecture might promote more rapid convergence. 

The necessary separation of inhibitory constraint functions was achieved by modifying equation \ref{eq:TotalSimpleConstraints} to:

\begin{equation}
I^{cons}_i =  g(\sum_{k=1}^{D_i} \beta^{1_D}  d_k ) (\sum_{k=1}^{C_i} I^C_k + \sum_{k=1}^{P_i} \gamma^{1_D} p_k )   
\label{eq:ExtInhibConstraint}
\end{equation}
\noindent where $I^C_k$ is non-specific contextual background excitation (see below), $D_i$ is the number of negative constraint cells synapsing on cell $i$, $P_i$ the number of positive constraint cells synapsing on cell $i$, $C_i$ the number contextual inputs (here, $C=1$ throughout), and the function $g(z)$ is an inverse sigmoid-type non-linearity of the form
\begin{equation}
g(z)=1-[\frac{1}{2} ( tanh(s (z-o))+1)]
\label{gZnonlin}
\end{equation}
\noindent where $s$ is the slope and $o$ the offset. Note that this function takes only values of $0...1$. To facilitate mathematical analysis, we use a linear approximation  $g(z)=1-min(max(sz,0),1)$. When this function is not in saturation it assumes value $sz$, where $s$ is the slope with $0<s<1$.      
      
In this extended version of the WTA ($WTA^E$), the inhibitory constraints enter as the argument of a non-linear function, $g$. This scales the effect of the two network excitatory sources (the positive constraint cells $p_k$ and the non-specific background excitation $I^C$).  Thus, each excitatory unit $x_i$ of the WTA$^E$ now receives two kinds of excitatory input: the forward input $I^{bias}_i$, and a constraint input $I^{cons}_i$. Note the critical difference: in $WTA^E$, $I^{cons}_i$ is always positive. In contrast, in the standard WTA this input may also be negative. This has the effect that negative constraints can never overwrite the forward input.
 
The dynamics of excitatory units in the extended WTA $x_i$ are:
 \begin{equation}
 \tau \dot{x}_i + G x_i = f_s( \alpha x_i - \beta^1 x_{N} - T_i   + I^{bias}_i + I^{cons}_i )
 \label{eq:ExtExcitConstraint}
 \end{equation}

Note that in Equation \ref{eq:ExtExcitConstraint}, the external input $I_i$ no longer appears because it is replaced by the contextual input $I^C$ that is now applied through the non-linearity $g(z)$. 

\subsection{Enhanced performance of $WTA^{e}$ networks}
We tested the performance of the extended WTA networks on all of the constraint satisfaction tasks. Unlike the standard networks, the $WTA^{e}$ networks solve all the problems presented. For graph coloring, the $WTA^{e}$ architecture converged significantly faster (Fig \ref{fig:graphcolor-all}G) and reduced the number of errors more rapidly (Fig \ref{fig:graphcolor-all}D,F). This difference became more apparent the larger the problem size. For example, networks with 49 nodes converged on average after $255\tau$. In contrast, the same problems required on average $1100\tau$ for $WTA^{s}$ (Fig \ref{fig:graphcolor-all}G).

Also, for MIS, we found that the $WTA^{e}$ architecture performed better for the larger problem sizes, with a speedup of up to 60\% (Fig \ref{fig:MIS-all}E,F).

The greatest advantage of the $WTA^{e}$ network was for SUD problems. $WTA^{e}$ solved all Sudoku problems quickly: convergence was on average $161\tau$ for the hard problem shown in Fig \ref{fig:sudoku-results}A and $142\tau$ for a set of 50 different sudoku standard problems of varying difficulty \cite{ProjectEuler}(Fig \ref{fig:sudoku-results}D-F). By contrast, the $WTA^{s}$ network was only able to solve 20\% of the instances of the hard problem (Fig \ref{fig:sudoku-results}D, gray) and 60\% of the 50 different problems in  (Fig \ref{fig:sudoku-results}F, gray) the same time in which $WTA^{e}$ solved 100\% of all problems. Note that because the solution times have a heavy tailed distribution, even for $WTA^{e}$ a small minority of runs take much longer. For example, whereas mean convergence is $142\tau$ for the 50 different problems, a few problems required up to 2000$\tau$ (Fig \ref{fig:sudoku-results}F).

\section{Discussion}

Our results indicate that large distributed constraint satisfaction problems can be processed on a computational substrate composed of many stereotypically connected neuronal WTA modules, together with a smaller number of more specific 'programming' neurons that interconnect the WTAs and thereby encode the constraints (or rules) of the particular problem to be solved. 

The architecture of these CSP networks is consistent with the strong recurrent excitatory, and recurrent inhibitory, connection motifs observed in the physiological and anatomical connections between neurons in superficial layers of mammalian neocortex \cite{douglas_intracellular_1991,binzegger_quantitative_2004}. Therefore our results are relevant to understanding how these biological circuits might operate. However, we do not consider that GC4P, MIS, and SUD are {\it per se} the problems solved by the actual neuronal cortical circuits. Instead, we choose these canonical CSP examples because their properties and applications are well understood in the computational literature, and so they can be used as a basis of comparison for the operation and performance of our WTAs networks, and then by extrapolation also of neocortical networks. 

Graph-coloring CSPs are intriguing computational problems because their structure requires simultaneous distributed satisfaction of constraints. However, in practice they are solved by sequential localized algorithms \cite{russell_artificial_2010, wu_review_2015}. For example, CSPs can be solved by exhaustive search, in which candidate solutions are systematically generated and tested for validity. However, this approach does not scale well with problem size \cite{kumar1992algorithms}. For our 49 node GC4P and SUD graph, this strategy would require up to approximately $10^{29}$ and $10^{22}$ configurations to be tested, respectively. Various heuristic algorithms can (but are not guaranteed to) improve performance beyond that obtainable by exhaustive search. By contrast, the neural network we present here solves CSPs efficiently without relying on domain-specific heuristics. However this performance would be a property of the physically realized network, and not of the algorithmic simulation of the model network that we are obliged to use here. For the moment our estimates of network performance are in terms of model time steps $\tau$ (eg Fig \ref{fig:graphcolor-all}F), which stand proxy for physical performance measurements.   

Previous approaches to solving CSPs using artificial neural networks \cite{wang1991solving, Habenschuss13,JonkeMaass16,HopfieldTank1985,Mostafa_etal13b,mezard2009constraint,mcclelland2014interactive,RosenfeldZucker1976,MillerZucker1999} have relied on the use of saturating neurons to maintain global stability, and have neglected the important role of instability. Output saturation is not observed in biological networks, where neurons typically operate at well beneath their maximum discharge rate. The computational implications of this well-recognized fact have until recently received little attention. For example, it is now becoming clear that non-saturating 'ReLu' activation functions are advantageous for deep learning networks \cite{nair2010rectified, maas2013rectifier,Lecun2015deep}. Here we now show that there are novel principles of network computation that depend on non-saturating activation. In this case stability relies on shared inhibition, which allows transient periods of highly unstable dynamics. This kind of instability does not exist in networks that utilize saturating neurons (eg \cite{HopfieldTank1985,MillerZucker1999}) where the majority (or even all) neurons are in either positive or negative saturation. In networks with non-saturating LTNs, the derivative of the activation function (equal to 1 here) appears in $\mathbf{J}_{eff}$ for all supra-threshold neurons, 
and hence all currently active neurons contribute to the expansion or contraction of the network dynamics. Neurons that are in saturation are of course also active, but because the derivative of their saturated activation function is at or near zero they contribute little or nothing to the expansion or contraction of the network dynamics. Analyzing the properties of $\mathbf{J}_{eff}$ is thus a powerful tool to understand the way by which our networks switch between different states autonomously as driven by their dynamics during the unstable parts of the dynamics. This tool is analogous to the energy function used in work pioneered by \cite{HopfieldTank1985}, analysis of which has provided great insight into how saturating networks compute.

Our approach consists of a set of rules that allows the systematic 'programming' of biologically plausible networks. Thus, we are able to program the desired computational processes onto a uniform substrate in a scalable manner \cite{rutishauser_computation_2015, RutishauserDouglas2011}. This approach has technological benefits for configuring large scale neuromorphic hardware such as IBM TrueNorth \cite{merolla2014million} or the Dynap chip \cite{qiao2015reconfigurable}, which instantiate a physical network rather than simulating it as we are doing here. While we deal here only with continously-valued networks, it has been shown that the types of WTA-networks we use here can also be implemented using spiking neurons \cite{Neftci2013,Neftci_etal11,wang2006fast}.

Although we do not claim that our CSP problems are implemented in real cortical networks, principles such as instability as the driving force of computation \cite{rutishauser_computation_2015}, and search through forbidden sets, are likely  fundamental to spontaneous computation in all types of LTN networks. In addition, this work provides insight into how analogous hardware should be engineered. These practical implications are in contrast to more general theoretical frameworks (e.g. \cite{heeger2017theory}) that often lack a circuit-level implementation and so cannot make predictions about the necessary computational roles of cell types such as we do here. Note also that the computational properties of the networks we describe here are preserved regardless of network size. This is because all aspects of the network rely on a simple computational motif (that of the WTA) that can be replicated as many times as needed for a particular problem without having to make modifications that depend on network size. This scalability is in contrast to other attractor-based computational approaches, which can be shown to solve small problems but cannot easily be generalized to larger ones \cite{afraimovich2004}.

\subsection{Computational properties of network solution of CSP}
The fundamental operation of our network involves simultaneous and interactive selection of values (states) across the WTA modules. The selection process drives the activities of some neurons below threshold using the signal gain developed by those neurons which receive support from either local or remote excitatory input. The dynamics of the network forbid partitions of active neurons that are inconsistent with network stability. These partitions are left exponentially quickly because the unstably high gain generated by the neurons active in the forbidden partition will increase recurrent inhibition such that at least one active will be driven beneath its activation threshold \cite{rutishauser_computation_2015}. As a consequence of this, a new partition of lower divergence is entered. This next partition may again be forbidden, and so exited; or it might be permitted, and so stable. Exploiting instability in this manner can be thought of as taking the path of least resistance in state space \cite{NiemeyerSlotine1997}. This exploration of state space continues until a consistent permitted partition is found \cite{Hahnloser00,rutishauser_computation_2015}. In the absence of noise or other external inputs, any transition between two states results in a reduction of divergence. This reduction implies that the network cannot return  to its previously occupied forbidden state, and so introduces a form of memory into the network that prevents cycling. In the presence of noise cycling becomes theoretically possible, but is very unlikely \cite{rutishauser_computation_2015}.

Solving CSPs by sequentially exploring different network subspaces has interesting similarities with algorithmic linear optimization methods, in particular the simplex and related methods \cite{MillerZucker1992,MillerZucker1999}. The critical step in simplex is the Pivot, an algorithmic manipulation that improves the subset of problem variable(s) to be maximized, while holding all others constant. This involves a decision, followed by a change of basis, and then maximization along the newly chosen dimensions. This process is similar to the process whereby the WTA network transitions through its forbidden sets (a link between neural network operation and simplex similar to that which has been made through the equivalence between polymatrix games and CSPs in \cite{MillerZucker1992} for relaxation labeling networks \cite{RosenfeldZucker1976} with saturating units).  Driven by the exponentially shrinking volumes implied by negative divergence, the unstable dynamics rapidly cause a switch to a different state of lower dimensionality of the state space. The direction in which the expansion proceeds is described by the subset of eigenvectors of the effective Jacobian $\mathbf{J}_{eff}$ that have positive eigenvalues \cite{rutishauser_computation_2015}. This network step is similar to maximization of the chosen variable in simplex. Furthermore, whenever the network switches from a forbidden set to another set (which is either forbidden or permitted), the network performs a ‘Pivot’, changing the basis functions among which the dynamics evolve. In contrast, the mechanism by which the CSP network implements these search principles is fundamentally different from linear programming and similar approaches, including the generalization to CSPs based on polymatrix games based on Lemke's algorithm \cite{MillerZucker1992}. Firstly, our network performs these steps fully asynchronously and autonomously for every module. Secondly, our network does not require access to a global cost function, does not require access to the current values of all variables, and does not depend on an external controller to decide suitable pivots. Instead, our network moves along the best directions for each WTA, and so the search proceeds for all WTAs in parallel. 

The constraint connections affect the mutual selection (value assignment) process of the coupled WTA variables. These constraints are implemented by directed weighted connections that provide an immediate and distributed $\epsilon$ update of all the appropriate values of all affected variables. Within each subspace, the network behaves as a piecewise linear system that is computationally powerful in that the partial derivatives of the system update the many interacting variables simultaneously and consistently as described by the effective Jacobian $\mathbf{J}_{eff}$. 

Importantly, these are updates to possible mixtures of values evolving at each WTA variable, rather than a replacement of one discrete value by another. But, eventually each WTA variable will enhance the signal due to one candidate value, while suppressing its competitors. This process is radically different to the back-tracking / constraint-propagation method implemented by digital algorithms that procedurally generate and test the consequences of alternative discrete value assignments to particular variables. Instead, the CSP network approaches its solution by successive approximation of candidate quality, and so is unlikely to compute towards a false assignment (unless the CSP has no feasible solution). This optimization process follows from the computational dynamics of the network: Its computational trajectory follows successively less unfavorable forbidden subspaces until a permitted subspace is entered.  

A further important distinction between algorithmic CS and our network rests in the assignment of an initial candidate configuration. Algorithmic approaches begin with a candidate configuration in which every variable is initialized with some legal value. By contrast, the initial CSP network assignment is effectively null: All $x_i$ at all WTA neurons are zero. However, these values are soon affected by the stochastic context signals $I^C$ applied to all $x_i$ so that there is almost immediately a low amplitude mixture  of variables across the network $x_i$.  The network dynamics then bootstrap better estimates of these mixtures through the constraints until the network finally converges towards a complete and consistent assignment. In this way the network offers an novel approach to CSP that is a dynamic balance between candidate generation and validation through progressive refinement of a mixture of values at each variable. 

\subsection{Probabilistic processing}
A certain degree of noise is essential for the operation of our networks. Such stochasticity and bias enters into the CSP network process via inputs from its embedding network. These are the contextual excitatory inputs $I_C$. For the moment, the $I_C$ introduce only randomness and biases that enable the CSP network to gain access to an otherwise computationally inaccessible solution subspace, and so provide some degree of innovation in the search process. It is also these inputs that are modulated by multiplicative inhibition. In a more realistic scenario, $I^C$ would be replaced with input from other parts of the brain to specify priors. This way, the network would be responsive to constraints set by other parts of the network, such as sensory input or internal states.

\begin{figure}
\centering
\includegraphics[angle=0,width=15cm]{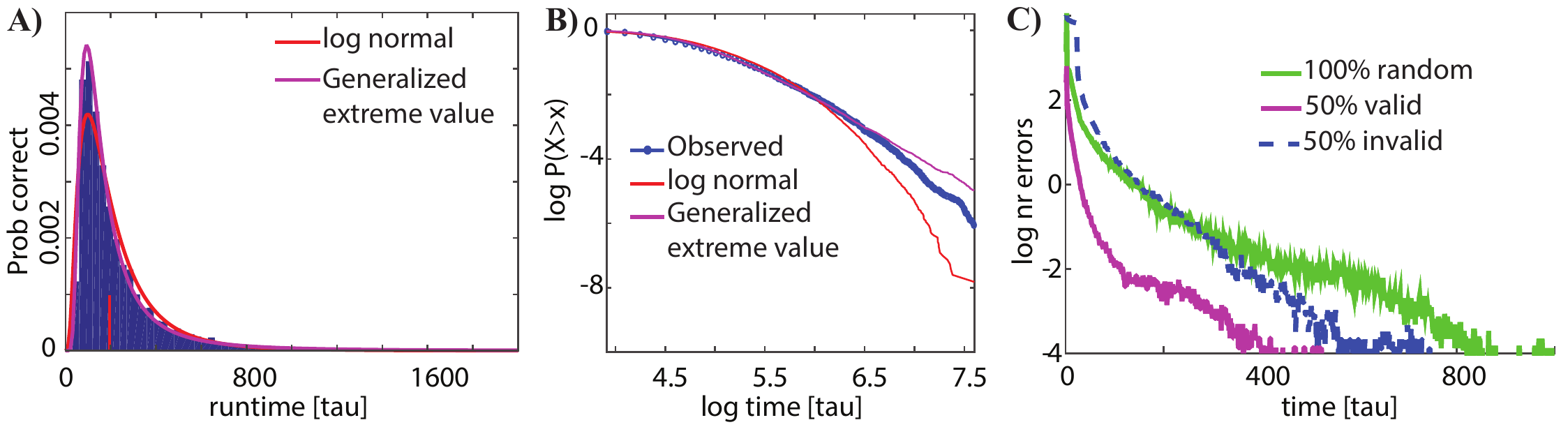} 
\caption{Distribution of run times and influence of initial conditions.
(A) Probability that a simulation will find a correct solution after a certain amount of simulation time. N=6000 simulation runs of random graphs with N=36 nodes, same parameters as in Fig \ref{fig:graphcolor-all}H). The majority of simulations find a solution within 200$\tau$ (red line).
The data was well fit by the log-normal ($\mu=5.11-5.15, \sigma=0.72-0.75$, 95\% confidence intervals) and generalized extreme value ($k=0.50-0.56, \sigma=74.3-78.7, \mu=120.5-125.0$, 95\% confidence intervals) distributions. Both these distributions are characteristic of heavy-tailed phenomena \protect\cite{Feldman1998practical}.
(B) Assessment of fit using a log-log plot. For robustness the y-axis is cumulative rather than log frequency. The tail of the observed data falls between the two theoretical distributions, indicating that its tail is heavier than expected by log-normal but less heavy than expected by generalized extreme value.
(C) GC4P for an identical N=25 node graph, but with different initial conditions: i) random initial conditions (green), ii) partially informative (50\% of states are
set correctly), and iii) partially uninformative (50\% of states are set incorrectly).
}
\label{fig:parallel-results}
\end{figure}

We confirmed that the network does find a solution more quickly when biased towards easy candidates (Fig. \ref{fig:parallel-results}C). Repetitions of these runs using the same random seed for the initial state (but not for further processing) confirm that non-deterministic processing nevertheless gives rise to distinct distributions of solution times for the easier and harder problems (Fig  \ref{fig:parallel-results}C). In the case of SUD, the prior information is set by the feedforward inputs $I^{bias}$ that implement the required values of some variables. The larger the number of biases, the harder the problem. But however hard, the set of problem biases for a given game of Sudoku are {\it a priori} known to be compatible with a solution, and therefore the network can be expected to finally find a complete solution. However, for more general coloring problems the input biases may only be desirable, and not known to be compatible with a complete assignment. In such cases the network may find an approximate but incomplete assignment (Fig \ref{fig:parallel-results}C).              

\subsection{Distribution of run times}
The network solves the majority of random graphs quickly, with a minority taking much longer (Fig \ref{fig:parallel-results}). Thus, the distribution of times required for the WTA networks to solve CSP is heavy-tailed (Fig \ref{fig:parallel-results}A,B). The form of the distribution does not depend on the hardness of the problem (compare Figs \ref{fig:sudoku-results}D and \ref{fig:sudoku-results}F). The heavy-tail persists even if the very same problem is solved multiple times from the same initial conditions, indicating that probabilistic processing allows the network to follow different trajectories that may differ substantially in their length (duration) because of alternative routes through successive forbidden subspaces \cite{rutishauser_computation_2015}. On the other hand if the network is seeded with initial conditions that favor simple solutions, then the median processing time is shorter than when the seed is biased towards invalid solutions (Fig  \ref{fig:parallel-results}C. Thus, the overall distribution of network run times appears to be a composition of the distribution over the hardness problems, as well the distributions over probabilistic trajectories. Such a composition is a characteristic of heavy-tailed processes \cite{Gomes2000heavy}.

\subsection{Computational advantage of multiplicative inhibition}
We found that the ability of the network to solve hard problems was improved significantly when negative constraints were implemented by non-linear multiplicative inhibition (see Fig \ref{fig:graphcolor-all}D,F). This non-linearity improved performance on difficult problems that have both global as well as local constraints, such as SUD. 

Multiple types of inhibition are a prominent feature of nervous systems \cite{blomfield_arithmetical_1974,koch_nonlinear_1983,Koch1998}, in particular in neocortex \cite{isaacson_how_2011}. There are both non-linear and linear mechanisms, associated with $GABA_A$ chloride-, and $GABA_B$ potassium-mediated inhibition, respectively.  Their actions are separable in intracellular recordings {\it in vivo} \cite{douglas_functional_1991,borg-graham_visual_1998}, but their effects during processing are mixed \cite{el-boustani_response-dependent_2014,zhang_nonlinear_2013}. 

Inhibition is mediated by distinct types of neuron encountered in the superficial layers of neocortex \cite{rudy_three_2011}. One large group (~40\%), the basket (BCs) inhibitory neurons, have horizontally disposed axons that target predominantly the soma and proximal dendritic segments of pyramidal neurons. The somatic bias of the synapses of these neurons make them likely candidates for implementing the somatic WTA selective mechanism. Another large group (~30\%), the bitufted  cells or double-bouquet cells (DBCs), have vertically disposed axons that target predominantly the more distal dendritic segments of pyramidal neurons. These neurons are candidates for the non-linear NC cells of our model. Although their particular conductance mechanisms are as yet unknown, non-linear inhibitory effects are considered to play an important role in the processing of synaptic inputs by dendrites of pyramidal cells \cite{Koch1998,bar-ilan_role_2013,brunel_single_2014,stuart_dendritic_2015}.

\begin{figure}
\centering
\includegraphics[angle=0,width=18cm]{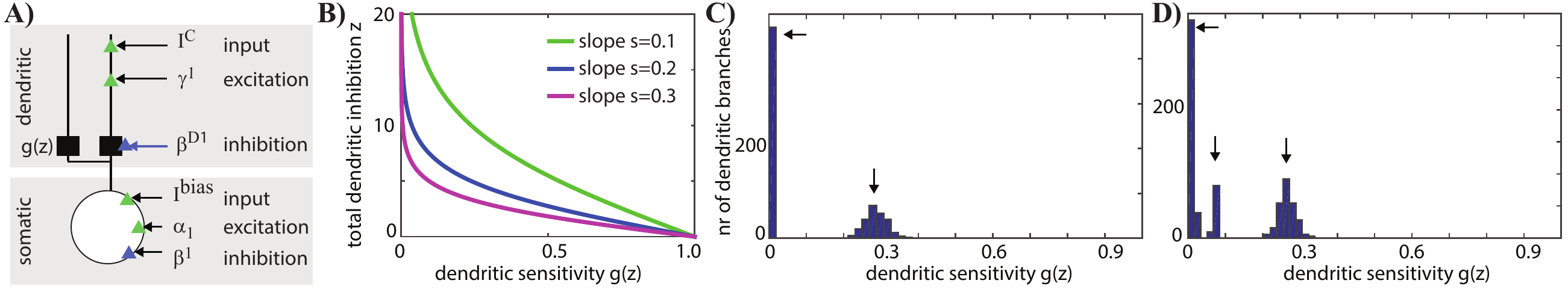} 
\caption{Behavior of dendritic non-linearity that improves performance of network in solving heavily constrained problems such as sudoku (SUD). 
(A) Separation of processing into a dendritic (top) and somatic (bottom) compartment by non-linearity g(z). Both compartments receive 
excitation and inhibition from nearby neurons as well as external inputs.
(B) Shape of the non-linearity g(z), where z is equal to the total inhibitory dendritic input that a dendritic branch receives. $g(z)=1-tanh(s z)$ is plotted for different values of s. The remainder of the fig uses $s=0.2$.
(C) Histogram of g(z) values across all dendritic branches in a simulation of sudoku (81 nodes) after a correct solution was found.
Note the bimodality (arrows): 64\% of all compartments have $g(z)=0$, making them insensitive to dendritic input. This is because their somatic
inputs $I_{bias}$ are strong. Effectively, this reduces the dimensionality of the problem. 
(D) Same as (C), but for a simulation where two different values of $I_{bias}$ were used (10 and 3). This results in a tri-modal distribution (arrowheads),
with the new mode corresponding to units with non-zero but weak $I_{bias}$. These units thus remain sensitive to dendritic input, but much less so
than the units where $I_{bias}=0$ (right-most mode).} 
\label{fig:shuntExplanation}
\end{figure}

When negative constraints are implemented by direct subtractive inhibition applied to the somatic compartment, they degrade the local WTA selection process by falsely contributing to the inhibitory normalization over the WTA $x_i$. We overcame this disadvantage by introducing a second, dendritic, compartment that receives the positive constraint and contextual input, and whose output $I^i_{dendrite}$ to the soma is governed by the non-linearity $g(z)$ (Fig \ref{fig:shuntExplanation}A). The somatic compartment  receives the standard WTA inputs $I^{bias}$, $I^{\alpha}$, and $I^{\beta1}$, and the output of the dendritic compartment. Thus, in addition to local recurrence, each excitatory unit $x_i$ of the WTA$^E$ receives two kinds of input: direct somatic input $I^{bias}_i$ and dendritic input $I^i_{dendrite}$. In this configuration,  multiplicative inhibitory constraints quench only the various sources of $I_C$ excitation received by each of the $x_i$, and do not interfere directly with the WTA local decision for the best supported of the $x_i$. This advantage explains the improved performance of the $WTA^E$ networks depicted in Figs \ref{fig:graphcolor-all}D,F,I; \ref{fig:MIS-all}D,E,F; and \ref{fig:sudoku-results}D,E,F.  

The $g(z)$ non-linearity provides 'on-path', multiplicative or 'shunting' inhibition previously described in biological dendrites  \cite{koch_nonlinear_1983,gidon_principles_2012}. This type of inhibition can veto excitatory input arriving on the the same dendritic branch but has much less influence on excitatory inputs that arrive on other branches or the soma \cite{zhang_nonlinear_2013}. The $g(z)$  has two important implications for computation. Firstly, somata that are strongly activated (e.g. by large $I^{bias}$) have little or no dendritic sensitivity because the strong feedback inhibition drives $g(z)$ towards $0$ (Fig \ref{fig:shuntExplanation}C). They ignore their dendritic excitatory input, thereby reducing the dimensionality of the problem. Second, dendritic sensitivity is graded so that when somatic activation varies, neurons with the low somatic activation are more sensitive to remote excitatory input (Fig \ref{fig:shuntExplanation}D). This mechanism provides a weighting of the importance of different dimensions of the problem, making it easier for solutions with little evidence to switch to alternative solutions, by comparison with those that have more evidence (ie more somatic activation).

The single dendritic compartment can be generalized to multiple compartments, each governed by its own nonlinearity, thereby allowing localized inhibitory modulation of specific excitatory input in the manner of a dendritic tree \cite{Koch1998,tran-van-minh_contribution_2015}. The total dendritic input $I^i_{dendrite}$ is then the sum of currents provided by all dendritic branches $j$. There are $\Delta^i$ branches in total. Each branch receives inhibitory inputs $d_k$ from negative constraint cells as well as two kinds of excitatory inputs: contextual inputs $I^k_C$ and inputs from the positive constraint cells $p_k$. Each branch $j$ receives $I_j$, $C_j$, and $P_j$ such inputs, respectively. 

\begin{equation}
I^i_{dendrite} = \sum_{j}^{\Delta^i} [ g(\sum_{k=1}^{I_j} \beta^D_1  d_k ) (\sum_{k=1}^{C_j} I^k_C + \sum_{k=1}^{P_j} \gamma^D_1 p_k )   ]
\label{eq:TotalEnhancedConstraints}
\end{equation}

Future work will explore the potential benefits for network processing of such parallel, or tree-structured, dendritic structures. For the present we consider only a single dendritic segment.  

\section{Conclusion}
We have shown that large distributed constraint satisfaction problems can be processed on a computational substrate that is composed of stereotypically connected neuronal WTA modules, and a smaller number of more specific 'programming' neurons which embody the constraints (or rules) of the particular problem to be solved. The rules of network construction and accompanying mathematical proofs guarantee that any instance of the three CSP types GC4P, MIS, and SUD implemented in the way described will find a solution. Note that the CSPs we considered can be reduced to graph coloring of planar (GC4P, MIS) and non-planar (SUD) graphs with the number of colors available given a-priori.

The networks use a combination of unstably high gain and network noise to drive a search for a consistent assignment values to problem variables. The organization of the network imposes constraints on the evolving manifold of system dynamics, with the result that the computational trajectory of the network is steered toward progressive satisfaction of all the problem constraints. This process takes advantage of the non-saturating nature of the individual neurons, which results in the effective Jacobian being driven by all neurons that are currently above threshold. 

This search performance is greatly improved if the mechanism of value selection at any variable can reduce its sensitivity to constraints according the confidence of selection. This can be achieved by using subtractive inhibition for selection, while modulating constraint inputs using multiplicative inhibition. This arrangement allows the constraint satisfaction network to solve more difficult problems, and to solve all such problems more quickly.

Our findings provide insight into the operation of the neuronal circuits of the neocortex, where the fundamental patterns of connection amongst superficial neurons is consistent with the WTA networks described here \cite{douglas_neuronal_2004,lee2016anatomy,rudy_three_2011,binzegger_quantitative_2004}. Our findings are also relevant to the design and construction of hybrid analog digital neuromorphic processing systems  \cite{liu_event-based_2015,Indiveri09,Neftci2013} because they provide general principles whereby a physical computational substrate could be engineered and utilized.

\section{Methods}
\subsection{Numerical simulations}
All simulations were implemented in MATLAB. Numerical integration of the ODEs is with Euler integration with $\delta=0.01$.

The Boost graph library \cite{BGL2002} and its MATLAB interface MatlabBGL \cite{Gleich2009} is used for graph-theoretical algorithms such as confirmation that graphs are planar, generation of random graphs, Chrobak-Payne Straight Line Drawing etc. The description of the network is generated automatically based on an XML file that specifies the graph. The XML file is in JFF format as used by JFLAP \cite{Rodger06}. 

\subsection{Translating a graph coloring problem into collection of WTAs}
The graph is decomposed into fully connected (all-to-all) non-overlapping subgraphs, for each of which a NC cell ("ring") is added (one for each color, so 4 for each ring). The sum of all $\beta^{D}_1j$ projecting to a particular pyramid cell $j$ is normalized to a constant equal to $\beta^{D}_j$.
External input to all pyramid cells (4 for each node) is normally distributed i.i.d. noise (currently $\mu=1$ and $\sigma=0.1$, i.e. 10\% of the mean).

\section{Acknowledgements}
We acknowledge the many contributions of Kevan Martin, as well as the inspiration of the Capo Caccia Workshops on Neuromorphic Engineering. We acknowledge funding by the McKnight Endowment for Neuroscience, the National Science Foundation Award 1554105, and the National Institute of Health Award R01MH110831. 

\bibliographystyle{apacite}
\bibliography{../bibsNetwork,bibs_converted}

\begin{thebibliography}{}

\bibitem [\protect \citeauthoryear {%
Abbott%
}{%
Abbott%
}{%
{\protect \APACyear {1994}}%
}]{%
Abbott94}
\APACinsertmetastar {%
Abbott94}%
\begin{APACrefauthors}%
Abbott, L.%
\end{APACrefauthors}%
\unskip\
\newblock
\APACrefYearMonthDay{1994}{}{}.
\newblock
{\BBOQ}\APACrefatitle {Decoding neuronal firing and modelling neural networks}
  {Decoding neuronal firing and modelling neural networks}.{\BBCQ}
\newblock
\APACjournalVolNumPages{Q. Rev. Biophys.}{27}{}{291-331}.
\PrintBackRefs{\CurrentBib}

\bibitem [\protect \citeauthoryear {%
Afek%
\ \protect \BOthers {.}}{%
Afek%
\ \protect \BOthers {.}}{%
{\protect \APACyear {2011}}%
}]{%
Afek2011}
\APACinsertmetastar {%
Afek2011}%
\begin{APACrefauthors}%
Afek, Y.%
, Alon, N.%
, Barad, O.%
, Hornstein, E.%
, Barkai, N.%
\BCBL {}\ \BBA {} Bar-Joseph, Z.%
\end{APACrefauthors}%
\unskip\
\newblock
\APACrefYearMonthDay{2011}{}{}.
\newblock
{\BBOQ}\APACrefatitle {A biological solution to a fundamental distributed
  computing problem} {A biological solution to a fundamental distributed
  computing problem}.{\BBCQ}
\newblock
\APACjournalVolNumPages{Science}{331}{6014}{183--185}.
\PrintBackRefs{\CurrentBib}

\bibitem [\protect \citeauthoryear {%
Afraimovich%
, Rabinovich%
\BCBL {}\ \BBA {} Varona%
}{%
Afraimovich%
\ \protect \BOthers {.}}{%
{\protect \APACyear {2004}}%
}]{%
afraimovich2004}
\APACinsertmetastar {%
afraimovich2004}%
\begin{APACrefauthors}%
Afraimovich, V\BPBI S.%
, Rabinovich, M\BPBI I.%
\BCBL {}\ \BBA {} Varona, P.%
\end{APACrefauthors}%
\unskip\
\newblock
\APACrefYearMonthDay{2004}{}{}.
\newblock
{\BBOQ}\APACrefatitle {Heteroclinic contours in neural ensembles and the
  winnerless competition principle} {Heteroclinic contours in neural ensembles
  and the winnerless competition principle}.{\BBCQ}
\newblock
\APACjournalVolNumPages{International Journal of Bifurcation and
  Chaos}{14}{04}{1195--1208}.
\PrintBackRefs{\CurrentBib}

\bibitem [\protect \citeauthoryear {%
Appel%
, Haken%
\BCBL {}\ \BBA {} Koch%
}{%
Appel%
\ \protect \BOthers {.}}{%
{\protect \APACyear {1977}}%
}]{%
AppHakKoc77}
\APACinsertmetastar {%
AppHakKoc77}%
\begin{APACrefauthors}%
Appel, K.%
, Haken, W.%
\BCBL {}\ \BBA {} Koch, J.%
\end{APACrefauthors}%
\unskip\
\newblock
\APACrefYearMonthDay{1977}{}{}.
\newblock
{\BBOQ}\APACrefatitle {Every planar map is four colorable} {Every planar map is
  four colorable}.{\BBCQ}
\newblock
\APACjournalVolNumPages{Illinois Journal of Mathematics}{21}{}{429-567}.
\PrintBackRefs{\CurrentBib}

\bibitem [\protect \citeauthoryear {%
Bar-Ilan%
, Gidon%
\BCBL {}\ \BBA {} Segev%
}{%
Bar-Ilan%
\ \protect \BOthers {.}}{%
{\protect \APACyear {2013}}%
}]{%
bar-ilan_role_2013}
\APACinsertmetastar {%
bar-ilan_role_2013}%
\begin{APACrefauthors}%
Bar-Ilan, L.%
, Gidon, A.%
\BCBL {}\ \BBA {} Segev, I.%
\end{APACrefauthors}%
\unskip\
\newblock
\APACrefYearMonthDay{2013}{}{}.
\newblock
{\BBOQ}\APACrefatitle {The role of dendritic inhibition in shaping the
  plasticity of excitatory synapses} {The role of dendritic inhibition in
  shaping the plasticity of excitatory synapses}.{\BBCQ}
\newblock
\APACjournalVolNumPages{Front Neural Circuits}{6}{}{}.
\PrintBackRefs{\CurrentBib}

\bibitem [\protect \citeauthoryear {%
Ben-Yishai%
, Bar-Or%
\BCBL {}\ \BBA {} Sompolinsky%
}{%
Ben-Yishai%
\ \protect \BOthers {.}}{%
{\protect \APACyear {1995}}%
}]{%
ben-yishai_theory_1995}
\APACinsertmetastar {%
ben-yishai_theory_1995}%
\begin{APACrefauthors}%
Ben-Yishai, R.%
, Bar-Or, R\BPBI L.%
\BCBL {}\ \BBA {} Sompolinsky, H.%
\end{APACrefauthors}%
\unskip\
\newblock
\APACrefYearMonthDay{1995}{}{}.
\newblock
{\BBOQ}\APACrefatitle {Theory of orientation tuning in visual cortex.} {Theory
  of orientation tuning in visual cortex.}{\BBCQ}
\newblock
\APACjournalVolNumPages{{PNAS}}{92}{9}{3844--3848}.
\PrintBackRefs{\CurrentBib}

\bibitem [\protect \citeauthoryear {%
Binzegger%
, Douglas%
\BCBL {}\ \BBA {} Martin%
}{%
Binzegger%
\ \protect \BOthers {.}}{%
{\protect \APACyear {2004}}%
}]{%
binzegger_quantitative_2004}
\APACinsertmetastar {%
binzegger_quantitative_2004}%
\begin{APACrefauthors}%
Binzegger, T.%
, Douglas, R\BPBI J.%
\BCBL {}\ \BBA {} Martin, K.%
\end{APACrefauthors}%
\unskip\
\newblock
\APACrefYearMonthDay{2004}{}{}.
\newblock
{\BBOQ}\APACrefatitle {A quantitative map of the circuit of cat primary visual
  cortex} {A quantitative map of the circuit of cat primary visual
  cortex}.{\BBCQ}
\newblock
\APACjournalVolNumPages{The Journal of Neuroscience}{24}{39}{8441--8453}.
\PrintBackRefs{\CurrentBib}

\bibitem [\protect \citeauthoryear {%
Blomfield%
}{%
Blomfield%
}{%
{\protect \APACyear {1974-03-29}}%
}]{%
blomfield_arithmetical_1974}
\APACinsertmetastar {%
blomfield_arithmetical_1974}%
\begin{APACrefauthors}%
Blomfield, S.%
\end{APACrefauthors}%
\unskip\
\newblock
\APACrefYearMonthDay{1974-03-29}{}{}.
\newblock
{\BBOQ}\APACrefatitle {Arithmetical operations performed by nerve cells}
  {Arithmetical operations performed by nerve cells}.{\BBCQ}
\newblock
\APACjournalVolNumPages{Brain Research}{69}{1}{115--124}.
\PrintBackRefs{\CurrentBib}

\bibitem [\protect \citeauthoryear {%
Borg-Graham%
, Monier%
\BCBL {}\ \BBA {} Fregnac%
}{%
Borg-Graham%
\ \protect \BOthers {.}}{%
{\protect \APACyear {1998}}%
}]{%
borg-graham_visual_1998}
\APACinsertmetastar {%
borg-graham_visual_1998}%
\begin{APACrefauthors}%
Borg-Graham, L\BPBI J.%
, Monier, C.%
\BCBL {}\ \BBA {} Fregnac, Y.%
\end{APACrefauthors}%
\unskip\
\newblock
\APACrefYearMonthDay{1998}{}{}.
\newblock
{\BBOQ}\APACrefatitle {Visual input evokes transient and strong shunting
  inhibition in visual cortical neurons} {Visual input evokes transient and
  strong shunting inhibition in visual cortical neurons}.{\BBCQ}
\newblock
\APACjournalVolNumPages{Nature}{393}{6683}{369}.
\PrintBackRefs{\CurrentBib}

\bibitem [\protect \citeauthoryear {%
Brunel%
, Hakim%
\BCBL {}\ \BBA {} Richardson%
}{%
Brunel%
\ \protect \BOthers {.}}{%
{\protect \APACyear {2014-04}}%
}]{%
brunel_single_2014}
\APACinsertmetastar {%
brunel_single_2014}%
\begin{APACrefauthors}%
Brunel, N.%
, Hakim, V.%
\BCBL {}\ \BBA {} Richardson, M\BPBI J.%
\end{APACrefauthors}%
\unskip\
\newblock
\APACrefYearMonthDay{2014-04}{}{}.
\newblock
{\BBOQ}\APACrefatitle {Single neuron dynamics and computation} {Single neuron
  dynamics and computation}.{\BBCQ}
\newblock
\APACjournalVolNumPages{Current Opinion in Neurobiology}{25}{}{149--155}.
\PrintBackRefs{\CurrentBib}

\bibitem [\protect \citeauthoryear {%
Carandini%
\ \BBA {} Heeger%
}{%
Carandini%
\ \BBA {} Heeger%
}{%
{\protect \APACyear {2012}}%
}]{%
carandini_normalization_2012}
\APACinsertmetastar {%
carandini_normalization_2012}%
\begin{APACrefauthors}%
Carandini, M.%
\BCBT {}\ \BBA {} Heeger, D\BPBI J.%
\end{APACrefauthors}%
\unskip\
\newblock
\APACrefYearMonthDay{2012}{}{}.
\newblock
{\BBOQ}\APACrefatitle {Normalization as a canonical neural computation}
  {Normalization as a canonical neural computation}.{\BBCQ}
\newblock
\APACjournalVolNumPages{Nat Rev Neurosci}{13}{1}{51--62}.
\PrintBackRefs{\CurrentBib}

\bibitem [\protect \citeauthoryear {%
Dayan%
}{%
Dayan%
}{%
{\protect \APACyear {2008}}%
}]{%
dayan2008simple}
\APACinsertmetastar {%
dayan2008simple}%
\begin{APACrefauthors}%
Dayan, P.%
\end{APACrefauthors}%
\unskip\
\newblock
\APACrefYearMonthDay{2008}{}{}.
\newblock
{\BBOQ}\APACrefatitle {Simple substrates for complex cognition} {Simple
  substrates for complex cognition}.{\BBCQ}
\newblock
\APACjournalVolNumPages{Frontiers in neuroscience}{2}{2}{255}.
\PrintBackRefs{\CurrentBib}

\bibitem [\protect \citeauthoryear {%
Dayan%
\ \BBA {} Abbott%
}{%
Dayan%
\ \BBA {} Abbott%
}{%
{\protect \APACyear {2001}}%
}]{%
AbbottDayan01}
\APACinsertmetastar {%
AbbottDayan01}%
\begin{APACrefauthors}%
Dayan, P.%
\BCBT {}\ \BBA {} Abbott, L.%
\end{APACrefauthors}%
\unskip\
\newblock
\APACrefYear{2001}.
\newblock
\APACrefbtitle {Theoretical Neuroscience} {Theoretical neuroscience}.
\newblock
\APACaddressPublisher{Cambridge MA}{MIT Press}.
\PrintBackRefs{\CurrentBib}

\bibitem [\protect \citeauthoryear {%
Douglas%
, Koch%
, Mahowald%
\BCBL {}\ \BBA {} Martin%
}{%
Douglas%
\ \protect \BOthers {.}}{%
{\protect \APACyear {1999}}%
}]{%
douglas_role_1999}
\APACinsertmetastar {%
douglas_role_1999}%
\begin{APACrefauthors}%
Douglas, R\BPBI J.%
, Koch, C.%
, Mahowald, M.%
\BCBL {}\ \BBA {} Martin, K.%
\end{APACrefauthors}%
\unskip\
\newblock
\APACrefYearMonthDay{1999}{}{}.
\newblock
{\BBOQ}\APACrefatitle {The role of recurrent excitation in neocortical
  circuits} {The role of recurrent excitation in neocortical circuits}.{\BBCQ}
\newblock
\BIn{} P.~Ulinski\ (\BED), \APACrefbtitle {Models of Cortical Circuits} {Models
  of cortical circuits}\ (\BPGS\ 251--282).
\newblock
\APACaddressPublisher{}{Springer}.
\PrintBackRefs{\CurrentBib}

\bibitem [\protect \citeauthoryear {%
Douglas%
\ \BBA {} Martin%
}{%
Douglas%
\ \BBA {} Martin%
}{%
{\protect \APACyear {1991}}%
}]{%
douglas_functional_1991}
\APACinsertmetastar {%
douglas_functional_1991}%
\begin{APACrefauthors}%
Douglas, R\BPBI J.%
\BCBT {}\ \BBA {} Martin, K.%
\end{APACrefauthors}%
\unskip\
\newblock
\APACrefYearMonthDay{1991}{}{}.
\newblock
{\BBOQ}\APACrefatitle {A functional microcircuit for cat visual cortex.} {A
  functional microcircuit for cat visual cortex.}{\BBCQ}
\newblock
\APACjournalVolNumPages{The Journal of Physiology}{440}{1}{735--769}.
\PrintBackRefs{\CurrentBib}

\bibitem [\protect \citeauthoryear {%
Douglas%
\ \BBA {} Martin%
}{%
Douglas%
\ \BBA {} Martin%
}{%
{\protect \APACyear {2004}}%
}]{%
douglas_neuronal_2004}
\APACinsertmetastar {%
douglas_neuronal_2004}%
\begin{APACrefauthors}%
Douglas, R\BPBI J.%
\BCBT {}\ \BBA {} Martin, K.%
\end{APACrefauthors}%
\unskip\
\newblock
\APACrefYearMonthDay{2004}{}{}.
\newblock
{\BBOQ}\APACrefatitle {Neuronal circuits of the neocortex} {Neuronal circuits
  of the neocortex}.{\BBCQ}
\newblock
\APACjournalVolNumPages{Annu. Rev. Neurosci.}{27}{}{419--451}.
\PrintBackRefs{\CurrentBib}

\bibitem [\protect \citeauthoryear {%
Douglas%
\ \BBA {} Martin%
}{%
Douglas%
\ \BBA {} Martin%
}{%
{\protect \APACyear {2007}}%
}]{%
Douglas2007_recurrent}
\APACinsertmetastar {%
Douglas2007_recurrent}%
\begin{APACrefauthors}%
Douglas, R\BPBI J.%
\BCBT {}\ \BBA {} Martin, K.%
\end{APACrefauthors}%
\unskip\
\newblock
\APACrefYearMonthDay{2007}{}{}.
\newblock
{\BBOQ}\APACrefatitle {Recurrent neuronal circuits in the neocortex.}
  {Recurrent neuronal circuits in the neocortex.}{\BBCQ}
\newblock
\APACjournalVolNumPages{Curr Biol}{17}{13}{R496--R500}.
\PrintBackRefs{\CurrentBib}

\bibitem [\protect \citeauthoryear {%
Douglas%
, Martin%
\BCBL {}\ \BBA {} Whitteridge%
}{%
Douglas%
\ \protect \BOthers {.}}{%
{\protect \APACyear {1989}}%
}]{%
douglas_canonical_1989}
\APACinsertmetastar {%
douglas_canonical_1989}%
\begin{APACrefauthors}%
Douglas, R\BPBI J.%
, Martin, K.%
\BCBL {}\ \BBA {} Whitteridge, D.%
\end{APACrefauthors}%
\unskip\
\newblock
\APACrefYearMonthDay{1989}{}{}.
\newblock
{\BBOQ}\APACrefatitle {A Canonical Microcircuit for Neocortex} {A canonical
  microcircuit for neocortex}.{\BBCQ}
\newblock
\APACjournalVolNumPages{Neural Computation}{1}{4}{480--488}.
\PrintBackRefs{\CurrentBib}

\bibitem [\protect \citeauthoryear {%
Douglas%
, Martin%
\BCBL {}\ \BBA {} Whitteridge%
}{%
Douglas%
\ \protect \BOthers {.}}{%
{\protect \APACyear {1991}}%
}]{%
douglas_intracellular_1991}
\APACinsertmetastar {%
douglas_intracellular_1991}%
\begin{APACrefauthors}%
Douglas, R\BPBI J.%
, Martin, K.%
\BCBL {}\ \BBA {} Whitteridge, D.%
\end{APACrefauthors}%
\unskip\
\newblock
\APACrefYearMonthDay{1991}{}{}.
\newblock
{\BBOQ}\APACrefatitle {An intracellular analysis of the visual responses of
  neurones in cat visual cortex.} {An intracellular analysis of the visual
  responses of neurones in cat visual cortex.}{\BBCQ}
\newblock
\APACjournalVolNumPages{J Physiol}{440}{1}{659--696}.
\PrintBackRefs{\CurrentBib}

\bibitem [\protect \citeauthoryear {%
El-Boustani%
\ \BBA {} Sur%
}{%
El-Boustani%
\ \BBA {} Sur%
}{%
{\protect \APACyear {2014}}%
}]{%
el-boustani_response-dependent_2014}
\APACinsertmetastar {%
el-boustani_response-dependent_2014}%
\begin{APACrefauthors}%
El-Boustani, S.%
\BCBT {}\ \BBA {} Sur, M.%
\end{APACrefauthors}%
\unskip\
\newblock
\APACrefYearMonthDay{2014}{}{}.
\newblock
{\BBOQ}\APACrefatitle {Response-dependent dynamics of cell-specific inhibition
  in cortical networks in vivo} {Response-dependent dynamics of cell-specific
  inhibition in cortical networks in vivo}.{\BBCQ}
\newblock
\APACjournalVolNumPages{Nat Commun}{5}{}{}.
\PrintBackRefs{\CurrentBib}

\bibitem [\protect \citeauthoryear {%
Ercsey-Ravasz%
\ \BBA {} Toroczkai%
}{%
Ercsey-Ravasz%
\ \BBA {} Toroczkai%
}{%
{\protect \APACyear {2012}}%
}]{%
Ercsey2012chaos}
\APACinsertmetastar {%
Ercsey2012chaos}%
\begin{APACrefauthors}%
Ercsey-Ravasz, M.%
\BCBT {}\ \BBA {} Toroczkai, Z.%
\end{APACrefauthors}%
\unskip\
\newblock
\APACrefYearMonthDay{2012}{}{}.
\newblock
{\BBOQ}\APACrefatitle {The chaos within Sudoku} {The chaos within
  sudoku}.{\BBCQ}
\newblock
\APACjournalVolNumPages{Scientific reports}{2}{}{}.
\PrintBackRefs{\CurrentBib}

\bibitem [\protect \citeauthoryear {%
Ermentrout%
}{%
Ermentrout%
}{%
{\protect \APACyear {1992}}%
}]{%
Ermentrout92}
\APACinsertmetastar {%
Ermentrout92}%
\begin{APACrefauthors}%
Ermentrout, B.%
\end{APACrefauthors}%
\unskip\
\newblock
\APACrefYearMonthDay{1992}{}{}.
\newblock
{\BBOQ}\APACrefatitle {Complex dynamics in winner-take-all neural nets with
  slow inhibition} {Complex dynamics in winner-take-all neural nets with slow
  inhibition}.{\BBCQ}
\newblock
\APACjournalVolNumPages{Neural Networks}{5}{3}{415-431}.
\PrintBackRefs{\CurrentBib}

\bibitem [\protect \citeauthoryear {%
Feldman%
\ \BBA {} Taqqu%
}{%
Feldman%
\ \BBA {} Taqqu%
}{%
{\protect \APACyear {1998}}%
}]{%
Feldman1998practical}
\APACinsertmetastar {%
Feldman1998practical}%
\begin{APACrefauthors}%
Feldman, R.%
\BCBT {}\ \BBA {} Taqqu, M.%
\end{APACrefauthors}%
\unskip\
\newblock
\APACrefYear{1998}.
\newblock
\APACrefbtitle {A practical guide to heavy tails: statistical techniques and
  applications} {A practical guide to heavy tails: statistical techniques and
  applications}.
\newblock
\APACaddressPublisher{}{Springer Science \& Business Media}.
\PrintBackRefs{\CurrentBib}

\bibitem [\protect \citeauthoryear {%
Ferster%
, Chung%
\BCBL {}\ \BBA {} Wheat%
}{%
Ferster%
\ \protect \BOthers {.}}{%
{\protect \APACyear {1996}}%
}]{%
ferster_orientation_1996}
\APACinsertmetastar {%
ferster_orientation_1996}%
\begin{APACrefauthors}%
Ferster, D.%
, Chung, S.%
\BCBL {}\ \BBA {} Wheat, H\BPBI S.%
\end{APACrefauthors}%
\unskip\
\newblock
\APACrefYearMonthDay{1996}{}{}.
\newblock
{\BBOQ}\APACrefatitle {Orientation selectivity of thalamic inputs to cat visual
  cortex} {Orientation selectivity of thalamic inputs to cat visual
  cortex}.{\BBCQ}
\newblock
\APACjournalVolNumPages{Nature}{380}{}{249--252}.
\PrintBackRefs{\CurrentBib}

\bibitem [\protect \citeauthoryear {%
Gidon%
\ \BBA {} Segev%
}{%
Gidon%
\ \BBA {} Segev%
}{%
{\protect \APACyear {2012}}%
}]{%
gidon_principles_2012}
\APACinsertmetastar {%
gidon_principles_2012}%
\begin{APACrefauthors}%
Gidon, A.%
\BCBT {}\ \BBA {} Segev, I.%
\end{APACrefauthors}%
\unskip\
\newblock
\APACrefYearMonthDay{2012}{}{}.
\newblock
{\BBOQ}\APACrefatitle {Principles Governing the Operation of Synaptic
  Inhibition in Dendrites} {Principles governing the operation of synaptic
  inhibition in dendrites}.{\BBCQ}
\newblock
\APACjournalVolNumPages{Neuron}{75}{2}{330--341}.
\PrintBackRefs{\CurrentBib}

\bibitem [\protect \citeauthoryear {%
Gleich%
}{%
Gleich%
}{%
{\protect \APACyear {2009}}%
}]{%
Gleich2009}
\APACinsertmetastar {%
Gleich2009}%
\begin{APACrefauthors}%
Gleich, D\BPBI F.%
\end{APACrefauthors}%
\unskip\
\newblock
\APACrefYear{2009}.
\unskip\
\newblock
\APACrefbtitle {Models and Algorithms for PageRank Sensitivity} {Models and
  algorithms for pagerank sensitivity}\ \APACtypeAddressSchool {\BUPhD}{}{}.
\unskip\
\newblock
\APACaddressSchool {}{Stanford University}.
\unskip\
\newblock
\APACrefnote{Chapter 7 on MatlabBGL}
\PrintBackRefs{\CurrentBib}

\bibitem [\protect \citeauthoryear {%
Gomes%
, Selman%
, Crato%
\BCBL {}\ \BBA {} Kautz%
}{%
Gomes%
\ \protect \BOthers {.}}{%
{\protect \APACyear {2000}}%
}]{%
Gomes2000heavy}
\APACinsertmetastar {%
Gomes2000heavy}%
\begin{APACrefauthors}%
Gomes, C\BPBI P.%
, Selman, B.%
, Crato, N.%
\BCBL {}\ \BBA {} Kautz, H.%
\end{APACrefauthors}%
\unskip\
\newblock
\APACrefYearMonthDay{2000}{}{}.
\newblock
{\BBOQ}\APACrefatitle {Heavy-tailed phenomena in satisfiability and constraint
  satisfaction problems} {Heavy-tailed phenomena in satisfiability and
  constraint satisfaction problems}.{\BBCQ}
\newblock
\APACjournalVolNumPages{Journal of automated reasoning}{24}{1-2}{67--100}.
\PrintBackRefs{\CurrentBib}

\bibitem [\protect \citeauthoryear {%
Habenschuss%
, Jonke%
\BCBL {}\ \BBA {} Maass%
}{%
Habenschuss%
\ \protect \BOthers {.}}{%
{\protect \APACyear {2013}}%
}]{%
Habenschuss13}
\APACinsertmetastar {%
Habenschuss13}%
\begin{APACrefauthors}%
Habenschuss, S.%
, Jonke, Z.%
\BCBL {}\ \BBA {} Maass, W.%
\end{APACrefauthors}%
\unskip\
\newblock
\APACrefYearMonthDay{2013}{}{}.
\newblock
{\BBOQ}\APACrefatitle {Stochastic Computations in Cortical Microcircuit Models}
  {Stochastic computations in cortical microcircuit models}.{\BBCQ}
\newblock
\APACjournalVolNumPages{PLoS Comput Biol}{9}{}{e1003311}.
\PrintBackRefs{\CurrentBib}

\bibitem [\protect \citeauthoryear {%
Hahnloser%
, Douglas%
, Mahowald%
\BCBL {}\ \BBA {} Hepp%
}{%
Hahnloser%
\ \protect \BOthers {.}}{%
{\protect \APACyear {1999}}%
}]{%
Hahnloser99}
\APACinsertmetastar {%
Hahnloser99}%
\begin{APACrefauthors}%
Hahnloser, R.%
, Douglas, R\BPBI J.%
, Mahowald, M.%
\BCBL {}\ \BBA {} Hepp, K.%
\end{APACrefauthors}%
\unskip\
\newblock
\APACrefYearMonthDay{1999}{}{}.
\newblock
{\BBOQ}\APACrefatitle {Feedback interactions between neuronal pointers and maps
  for attentional processing} {Feedback interactions between neuronal pointers
  and maps for attentional processing}.{\BBCQ}
\newblock
\APACjournalVolNumPages{Nat Neurosci}{2}{8}{746-52}.
\PrintBackRefs{\CurrentBib}

\bibitem [\protect \citeauthoryear {%
Hahnloser%
, Sarpeshkar%
, Mahowald%
, Douglas%
\BCBL {}\ \BBA {} Seung%
}{%
Hahnloser%
\ \protect \BOthers {.}}{%
{\protect \APACyear {2000}}%
}]{%
Hahnloser00}
\APACinsertmetastar {%
Hahnloser00}%
\begin{APACrefauthors}%
Hahnloser, R.%
, Sarpeshkar, R.%
, Mahowald, M.%
, Douglas, R\BPBI J.%
\BCBL {}\ \BBA {} Seung, H\BPBI S.%
\end{APACrefauthors}%
\unskip\
\newblock
\APACrefYearMonthDay{2000}{}{}.
\newblock
{\BBOQ}\APACrefatitle {Digital selection and analogue amplification coexist in
  a cortex-inspired silicon circuit} {Digital selection and analogue
  amplification coexist in a cortex-inspired silicon circuit}.{\BBCQ}
\newblock
\APACjournalVolNumPages{Nature}{405}{6789}{947-51}.
\PrintBackRefs{\CurrentBib}

\bibitem [\protect \citeauthoryear {%
Heeger%
}{%
Heeger%
}{%
{\protect \APACyear {2017}}%
}]{%
heeger2017theory}
\APACinsertmetastar {%
heeger2017theory}%
\begin{APACrefauthors}%
Heeger, D\BPBI J.%
\end{APACrefauthors}%
\unskip\
\newblock
\APACrefYearMonthDay{2017}{}{}.
\newblock
{\BBOQ}\APACrefatitle {Theory of cortical function} {Theory of cortical
  function}.{\BBCQ}
\newblock
\APACjournalVolNumPages{Proceedings of the National Academy of
  Sciences}{}{}{201619788}.
\PrintBackRefs{\CurrentBib}

\bibitem [\protect \citeauthoryear {%
Hopfield%
}{%
Hopfield%
}{%
{\protect \APACyear {1982}}%
}]{%
Hopfield82}
\APACinsertmetastar {%
Hopfield82}%
\begin{APACrefauthors}%
Hopfield, J.%
\end{APACrefauthors}%
\unskip\
\newblock
\APACrefYearMonthDay{1982}{}{}.
\newblock
{\BBOQ}\APACrefatitle {Neural networks and physical systems with emergent
  collective computational abilities} {Neural networks and physical systems
  with emergent collective computational abilities}.{\BBCQ}
\newblock
\APACjournalVolNumPages{Proceedings of the National Academy of Sciences of the
  United States of America}{79}{}{2554-2558}.
\PrintBackRefs{\CurrentBib}

\bibitem [\protect \citeauthoryear {%
Hopfield%
\ \BBA {} Tank%
}{%
Hopfield%
\ \BBA {} Tank%
}{%
{\protect \APACyear {1985}}%
}]{%
HopfieldTank1985}
\APACinsertmetastar {%
HopfieldTank1985}%
\begin{APACrefauthors}%
Hopfield, J.%
\BCBT {}\ \BBA {} Tank, D\BPBI W.%
\end{APACrefauthors}%
\unskip\
\newblock
\APACrefYearMonthDay{1985}{}{}.
\newblock
{\BBOQ}\APACrefatitle {Neural computation of decisions in optimization
  problems} {Neural computation of decisions in optimization problems}.{\BBCQ}
\newblock
\APACjournalVolNumPages{Biological cybernetics}{52}{3}{141--152}.
\PrintBackRefs{\CurrentBib}

\bibitem [\protect \citeauthoryear {%
Horn%
}{%
Horn%
}{%
{\protect \APACyear {1985}}%
}]{%
Horn85}
\APACinsertmetastar {%
Horn85}%
\begin{APACrefauthors}%
Horn, R.%
\end{APACrefauthors}%
\unskip\
\newblock
\APACrefYear{1985}.
\newblock
\APACrefbtitle {Matrix analysis} {Matrix analysis}.
\newblock
\APACaddressPublisher{}{Cambridge University Press}.
\PrintBackRefs{\CurrentBib}

\bibitem [\protect \citeauthoryear {%
Indiveri%
, Chicca%
\BCBL {}\ \BBA {} Douglas%
}{%
Indiveri%
\ \protect \BOthers {.}}{%
{\protect \APACyear {2009}}%
}]{%
Indiveri09}
\APACinsertmetastar {%
Indiveri09}%
\begin{APACrefauthors}%
Indiveri, G.%
, Chicca, E.%
\BCBL {}\ \BBA {} Douglas, R\BPBI J.%
\end{APACrefauthors}%
\unskip\
\newblock
\APACrefYearMonthDay{2009}{}{}.
\newblock
{\BBOQ}\APACrefatitle {Artificial Cognitive Systems: From VLSI Networks of
  Spiking Neurons to Neuromorphic Cognition} {Artificial cognitive systems:
  From vlsi networks of spiking neurons to neuromorphic cognition}.{\BBCQ}
\newblock
\APACjournalVolNumPages{Cognitive Computation}{1}{2}{119-127}.
\PrintBackRefs{\CurrentBib}

\bibitem [\protect \citeauthoryear {%
Isaacson%
\ \BBA {} Scanziani%
}{%
Isaacson%
\ \BBA {} Scanziani%
}{%
{\protect \APACyear {2011}}%
}]{%
isaacson_how_2011}
\APACinsertmetastar {%
isaacson_how_2011}%
\begin{APACrefauthors}%
Isaacson, J.%
\BCBT {}\ \BBA {} Scanziani, M.%
\end{APACrefauthors}%
\unskip\
\newblock
\APACrefYearMonthDay{2011}{}{}.
\newblock
{\BBOQ}\APACrefatitle {How Inhibition Shapes Cortical Activity} {How inhibition
  shapes cortical activity}.{\BBCQ}
\newblock
\APACjournalVolNumPages{Neuron}{72}{2}{231--243}.
\PrintBackRefs{\CurrentBib}

\bibitem [\protect \citeauthoryear {%
Jiang%
, Wang%
, Lee%
, Stornetta%
\BCBL {}\ \BBA {} Zhu%
}{%
Jiang%
\ \protect \BOthers {.}}{%
{\protect \APACyear {2013}}%
}]{%
Jiang2013organization}
\APACinsertmetastar {%
Jiang2013organization}%
\begin{APACrefauthors}%
Jiang, X.%
, Wang, G.%
, Lee, A\BPBI J.%
, Stornetta, R\BPBI L.%
\BCBL {}\ \BBA {} Zhu, J\BPBI J.%
\end{APACrefauthors}%
\unskip\
\newblock
\APACrefYearMonthDay{2013}{}{}.
\newblock
{\BBOQ}\APACrefatitle {The organization of two new cortical interneuronal
  circuits} {The organization of two new cortical interneuronal
  circuits}.{\BBCQ}
\newblock
\APACjournalVolNumPages{Nature neuroscience}{16}{2}{210--218}.
\PrintBackRefs{\CurrentBib}

\bibitem [\protect \citeauthoryear {%
Jonke%
, Habenschuss%
\BCBL {}\ \BBA {} Maass%
}{%
Jonke%
\ \protect \BOthers {.}}{%
{\protect \APACyear {2016}}%
}]{%
JonkeMaass16}
\APACinsertmetastar {%
JonkeMaass16}%
\begin{APACrefauthors}%
Jonke, Z.%
, Habenschuss, S.%
\BCBL {}\ \BBA {} Maass, W.%
\end{APACrefauthors}%
\unskip\
\newblock
\APACrefYearMonthDay{2016}{}{}.
\newblock
{\BBOQ}\APACrefatitle {Solving constraint satisfaction problems with networks
  of spiking neurons} {Solving constraint satisfaction problems with networks
  of spiking neurons}.{\BBCQ}
\newblock
\APACjournalVolNumPages{Front. Neurosci., 30 March}{}{}{}.
\PrintBackRefs{\CurrentBib}

\bibitem [\protect \citeauthoryear {%
Kami{\'n}ski%
\ \protect \BOthers {.}}{%
Kami{\'n}ski%
\ \protect \BOthers {.}}{%
{\protect \APACyear {2017}}%
}]{%
Kaminski2017}
\APACinsertmetastar {%
Kaminski2017}%
\begin{APACrefauthors}%
Kami{\'n}ski, J.%
, Sullivan, S.%
, Chung, J\BPBI M.%
, Ross, I\BPBI B.%
, Mamelak, A\BPBI N.%
\BCBL {}\ \BBA {} Rutishauser, U.%
\end{APACrefauthors}%
\unskip\
\newblock
\APACrefYearMonthDay{2017}{}{}.
\newblock
{\BBOQ}\APACrefatitle {Persistently active neurons in human medial frontal and
  medial temporal lobe support working memory} {Persistently active neurons in
  human medial frontal and medial temporal lobe support working memory}.{\BBCQ}
\newblock
\APACjournalVolNumPages{Nature Neuroscience}{}{}{}.
\PrintBackRefs{\CurrentBib}

\bibitem [\protect \citeauthoryear {%
Karp%
}{%
Karp%
}{%
{\protect \APACyear {1972}}%
}]{%
Kar72}
\APACinsertmetastar {%
Kar72}%
\begin{APACrefauthors}%
Karp, R\BPBI M.%
\end{APACrefauthors}%
\unskip\
\newblock
\APACrefYearMonthDay{1972}{}{}.
\newblock
{\BBOQ}\APACrefatitle {Reducibility among combinatorial problems} {Reducibility
  among combinatorial problems}.{\BBCQ}
\newblock
\BIn{} R\BPBI E.~Miller\ \BBA {} J\BPBI W.~Thatcher\ (\BEDS), \APACrefbtitle
  {Complexity of Computer Computations} {Complexity of computer computations}\
  (\BPGS\ 85--103).
\newblock
\APACaddressPublisher{New York, USA}{Plenum Press}.
\PrintBackRefs{\CurrentBib}

\bibitem [\protect \citeauthoryear {%
Koch%
}{%
Koch%
}{%
{\protect \APACyear {1998}}%
}]{%
Koch1998}
\APACinsertmetastar {%
Koch1998}%
\begin{APACrefauthors}%
Koch, C.%
\end{APACrefauthors}%
\unskip\
\newblock
\APACrefYear{1998}.
\newblock
\APACrefbtitle {{Biophysics of Computation: Information Processing in Single
  Neurons (Computational Neuroscience)}} {{Biophysics of Computation:
  Information Processing in Single Neurons (Computational Neuroscience)}}.
\newblock
\APACaddressPublisher{}{Oxford University Press}.
\PrintBackRefs{\CurrentBib}

\bibitem [\protect \citeauthoryear {%
Koch%
, Poggio%
\BCBL {}\ \BBA {} Torre%
}{%
Koch%
\ \protect \BOthers {.}}{%
{\protect \APACyear {1983}}%
}]{%
koch_nonlinear_1983}
\APACinsertmetastar {%
koch_nonlinear_1983}%
\begin{APACrefauthors}%
Koch, C.%
, Poggio, T.%
\BCBL {}\ \BBA {} Torre, V.%
\end{APACrefauthors}%
\unskip\
\newblock
\APACrefYearMonthDay{1983}{}{}.
\newblock
{\BBOQ}\APACrefatitle {Nonlinear interactions in a dendritic tree:
  localization, timing, and role in information processing} {Nonlinear
  interactions in a dendritic tree: localization, timing, and role in
  information processing}.{\BBCQ}
\newblock
\APACjournalVolNumPages{Proceedings of the National Academy of
  Sciences}{80}{9}{2799--2802}.
\PrintBackRefs{\CurrentBib}

\bibitem [\protect \citeauthoryear {%
Koechlin%
\ \BBA {} Summerfield%
}{%
Koechlin%
\ \BBA {} Summerfield%
}{%
{\protect \APACyear {2007}}%
}]{%
koechlin2007}
\APACinsertmetastar {%
koechlin2007}%
\begin{APACrefauthors}%
Koechlin, E.%
\BCBT {}\ \BBA {} Summerfield, C.%
\end{APACrefauthors}%
\unskip\
\newblock
\APACrefYearMonthDay{2007}{}{}.
\newblock
{\BBOQ}\APACrefatitle {An information theoretical approach to prefrontal
  executive function} {An information theoretical approach to prefrontal
  executive function}.{\BBCQ}
\newblock
\APACjournalVolNumPages{Trends in cognitive sciences}{11}{6}{229--235}.
\PrintBackRefs{\CurrentBib}

\bibitem [\protect \citeauthoryear {%
Kumar%
}{%
Kumar%
}{%
{\protect \APACyear {1992}}%
}]{%
kumar1992algorithms}
\APACinsertmetastar {%
kumar1992algorithms}%
\begin{APACrefauthors}%
Kumar, V.%
\end{APACrefauthors}%
\unskip\
\newblock
\APACrefYearMonthDay{1992}{}{}.
\newblock
{\BBOQ}\APACrefatitle {Algorithms for constraint-satisfaction problems: A
  survey} {Algorithms for constraint-satisfaction problems: A survey}.{\BBCQ}
\newblock
\APACjournalVolNumPages{AI magazine}{13}{1}{32}.
\PrintBackRefs{\CurrentBib}

\bibitem [\protect \citeauthoryear {%
LeCun%
, Bengio%
\BCBL {}\ \BBA {} Hinton%
}{%
LeCun%
\ \protect \BOthers {.}}{%
{\protect \APACyear {2015}}%
}]{%
Lecun2015deep}
\APACinsertmetastar {%
Lecun2015deep}%
\begin{APACrefauthors}%
LeCun, Y.%
, Bengio, Y.%
\BCBL {}\ \BBA {} Hinton, G.%
\end{APACrefauthors}%
\unskip\
\newblock
\APACrefYearMonthDay{2015}{}{}.
\newblock
{\BBOQ}\APACrefatitle {Deep learning} {Deep learning}.{\BBCQ}
\newblock
\APACjournalVolNumPages{Nature}{521}{7553}{436--444}.
\PrintBackRefs{\CurrentBib}

\bibitem [\protect \citeauthoryear {%
Lee%
\ \protect \BOthers {.}}{%
Lee%
\ \protect \BOthers {.}}{%
{\protect \APACyear {2016}}%
}]{%
lee2016anatomy}
\APACinsertmetastar {%
lee2016anatomy}%
\begin{APACrefauthors}%
Lee, W\BHBI C\BPBI A.%
, Bonin, V.%
, Reed, M.%
, Graham, B\BPBI J.%
, Hood, G.%
, Glattfelder, K.%
\BCBL {}\ \BBA {} Reid, R\BPBI C.%
\end{APACrefauthors}%
\unskip\
\newblock
\APACrefYearMonthDay{2016}{}{}.
\newblock
{\BBOQ}\APACrefatitle {Anatomy and function of an excitatory network in the
  visual cortex} {Anatomy and function of an excitatory network in the visual
  cortex}.{\BBCQ}
\newblock
\APACjournalVolNumPages{Nature}{532}{7599}{370--374}.
\PrintBackRefs{\CurrentBib}

\bibitem [\protect \citeauthoryear {%
Li%
, Ibrahim%
, Liu%
, Zhang%
\BCBL {}\ \BBA {} Tao%
}{%
Li%
\ \protect \BOthers {.}}{%
{\protect \APACyear {2013}}%
}]{%
li_linear_2013}
\APACinsertmetastar {%
li_linear_2013}%
\begin{APACrefauthors}%
Li, Y\BHBI t.%
, Ibrahim, L\BPBI A.%
, Liu, B\BHBI h.%
, Zhang, L\BPBI I.%
\BCBL {}\ \BBA {} Tao, H\BPBI W.%
\end{APACrefauthors}%
\unskip\
\newblock
\APACrefYearMonthDay{2013}{}{}.
\newblock
{\BBOQ}\APACrefatitle {Linear transformation of thalamocortical input by
  intracortical excitation} {Linear transformation of thalamocortical input by
  intracortical excitation}.{\BBCQ}
\newblock
\APACjournalVolNumPages{Nat Neurosci}{16}{9}{1324--1330}.
\PrintBackRefs{\CurrentBib}

\bibitem [\protect \citeauthoryear {%
Lien%
\ \BBA {} Scanziani%
}{%
Lien%
\ \BBA {} Scanziani%
}{%
{\protect \APACyear {2013}}%
}]{%
lien_tuned_2013}
\APACinsertmetastar {%
lien_tuned_2013}%
\begin{APACrefauthors}%
Lien, A\BPBI D.%
\BCBT {}\ \BBA {} Scanziani, M.%
\end{APACrefauthors}%
\unskip\
\newblock
\APACrefYearMonthDay{2013}{}{}.
\newblock
{\BBOQ}\APACrefatitle {Tuned thalamic excitation is amplified by visual
  cortical circuits} {Tuned thalamic excitation is amplified by visual cortical
  circuits}.{\BBCQ}
\newblock
\APACjournalVolNumPages{Nat Neurosci}{16}{9}{1315--1323}.
\PrintBackRefs{\CurrentBib}

\bibitem [\protect \citeauthoryear {%
Liu%
, Delbruck%
, Indiveri%
, Whatley%
\BCBL {}\ \BBA {} Douglas%
}{%
Liu%
\ \protect \BOthers {.}}{%
{\protect \APACyear {2015}}%
}]{%
liu_event-based_2015}
\APACinsertmetastar {%
liu_event-based_2015}%
\begin{APACrefauthors}%
Liu, S\BHBI C.%
, Delbruck, T.%
, Indiveri, G.%
, Whatley, A.%
\BCBL {}\ \BBA {} Douglas, R\BPBI J.%
\end{APACrefauthors}%
\unskip\
\newblock
\APACrefYear{2015}.
\newblock
\APACrefbtitle {Event-based Neuromorphic Systems} {Event-based neuromorphic
  systems}.
\newblock
\APACaddressPublisher{}{John Wiley \& Sons, Ltd}.
\PrintBackRefs{\CurrentBib}

\bibitem [\protect \citeauthoryear {%
Lohmiller%
\ \BBA {} Slotine%
}{%
Lohmiller%
\ \BBA {} Slotine%
}{%
{\protect \APACyear {2000}}%
}]{%
Lohmiller2000}
\APACinsertmetastar {%
Lohmiller2000}%
\begin{APACrefauthors}%
Lohmiller, W.%
\BCBT {}\ \BBA {} Slotine, J.%
\end{APACrefauthors}%
\unskip\
\newblock
\APACrefYearMonthDay{2000}{}{}.
\newblock
{\BBOQ}\APACrefatitle {Nonlinear process control using contraction theory}
  {Nonlinear process control using contraction theory}.{\BBCQ}
\newblock
\APACjournalVolNumPages{AIChE Journal}{46}{3}{588-596}.
\PrintBackRefs{\CurrentBib}

\bibitem [\protect \citeauthoryear {%
Lynch%
}{%
Lynch%
}{%
{\protect \APACyear {1996}}%
}]{%
Lynch1996_distributed}
\APACinsertmetastar {%
Lynch1996_distributed}%
\begin{APACrefauthors}%
Lynch, N\BPBI A.%
\end{APACrefauthors}%
\unskip\
\newblock
\APACrefYear{1996}.
\newblock
\APACrefbtitle {Distributed algorithms} {Distributed algorithms}.
\newblock
\APACaddressPublisher{}{Morgan Kaufmann}.
\PrintBackRefs{\CurrentBib}

\bibitem [\protect \citeauthoryear {%
Maas%
, Hannun%
\BCBL {}\ \BBA {} Ng%
}{%
Maas%
\ \protect \BOthers {.}}{%
{\protect \APACyear {2013}}%
}]{%
maas2013rectifier}
\APACinsertmetastar {%
maas2013rectifier}%
\begin{APACrefauthors}%
Maas, A\BPBI L.%
, Hannun, A\BPBI Y.%
\BCBL {}\ \BBA {} Ng, A\BPBI Y.%
\end{APACrefauthors}%
\unskip\
\newblock
\APACrefYearMonthDay{2013}{}{}.
\newblock
{\BBOQ}\APACrefatitle {Rectifier nonlinearities improve neural network acoustic
  models} {Rectifier nonlinearities improve neural network acoustic
  models}.{\BBCQ}
\newblock
\BIn{} \APACrefbtitle {Proc. ICML} {Proc. icml}\ (\BVOL~30).
\PrintBackRefs{\CurrentBib}

\bibitem [\protect \citeauthoryear {%
Maass%
}{%
Maass%
}{%
{\protect \APACyear {2000}}%
}]{%
Maass00}
\APACinsertmetastar {%
Maass00}%
\begin{APACrefauthors}%
Maass, W.%
\end{APACrefauthors}%
\unskip\
\newblock
\APACrefYearMonthDay{2000}{}{}.
\newblock
{\BBOQ}\APACrefatitle {On the computational power of winner-take-all} {On the
  computational power of winner-take-all}.{\BBCQ}
\newblock
\APACjournalVolNumPages{Neural Computation}{12}{}{2519-2536}.
\PrintBackRefs{\CurrentBib}

\bibitem [\protect \citeauthoryear {%
McClelland%
, Mirman%
, Bolger%
\BCBL {}\ \BBA {} Khaitan%
}{%
McClelland%
\ \protect \BOthers {.}}{%
{\protect \APACyear {2014}}%
}]{%
mcclelland2014interactive}
\APACinsertmetastar {%
mcclelland2014interactive}%
\begin{APACrefauthors}%
McClelland, J\BPBI L.%
, Mirman, D.%
, Bolger, D\BPBI J.%
\BCBL {}\ \BBA {} Khaitan, P.%
\end{APACrefauthors}%
\unskip\
\newblock
\APACrefYearMonthDay{2014}{}{}.
\newblock
{\BBOQ}\APACrefatitle {Interactive activation and mutual constraint
  satisfaction in perception and cognition} {Interactive activation and mutual
  constraint satisfaction in perception and cognition}.{\BBCQ}
\newblock
\APACjournalVolNumPages{Cognitive science}{38}{6}{1139--1189}.
\PrintBackRefs{\CurrentBib}

\bibitem [\protect \citeauthoryear {%
Merolla%
\ \protect \BOthers {.}}{%
Merolla%
\ \protect \BOthers {.}}{%
{\protect \APACyear {2014}}%
}]{%
merolla2014million}
\APACinsertmetastar {%
merolla2014million}%
\begin{APACrefauthors}%
Merolla, P\BPBI A.%
, Arthur, J\BPBI V.%
, Alvarez-Icaza, R.%
, Cassidy, A\BPBI S.%
, Sawada, J.%
, Akopyan, F.%
\BDBL {}others%
\end{APACrefauthors}%
\unskip\
\newblock
\APACrefYearMonthDay{2014}{}{}.
\newblock
{\BBOQ}\APACrefatitle {A million spiking-neuron integrated circuit with a
  scalable communication network and interface} {A million spiking-neuron
  integrated circuit with a scalable communication network and
  interface}.{\BBCQ}
\newblock
\APACjournalVolNumPages{Science}{345}{6197}{668--673}.
\PrintBackRefs{\CurrentBib}

\bibitem [\protect \citeauthoryear {%
Mezard%
\ \BBA {} Mora%
}{%
Mezard%
\ \BBA {} Mora%
}{%
{\protect \APACyear {2009}}%
}]{%
mezard2009constraint}
\APACinsertmetastar {%
mezard2009constraint}%
\begin{APACrefauthors}%
Mezard, M.%
\BCBT {}\ \BBA {} Mora, T.%
\end{APACrefauthors}%
\unskip\
\newblock
\APACrefYearMonthDay{2009}{}{}.
\newblock
{\BBOQ}\APACrefatitle {Constraint satisfaction problems and neural networks: A
  statistical physics perspective} {Constraint satisfaction problems and neural
  networks: A statistical physics perspective}.{\BBCQ}
\newblock
\APACjournalVolNumPages{Journal of Physiology-Paris}{103}{1}{107--113}.
\PrintBackRefs{\CurrentBib}

\bibitem [\protect \citeauthoryear {%
Miller%
\ \BBA {} Zucker%
}{%
Miller%
\ \BBA {} Zucker%
}{%
{\protect \APACyear {1992}}%
}]{%
MillerZucker1992}
\APACinsertmetastar {%
MillerZucker1992}%
\begin{APACrefauthors}%
Miller, D\BPBI A.%
\BCBT {}\ \BBA {} Zucker, S\BPBI W.%
\end{APACrefauthors}%
\unskip\
\newblock
\APACrefYearMonthDay{1992}{}{}.
\newblock
{\BBOQ}\APACrefatitle {Efficient simplex-like methods for equilibria of
  nonsymmetric analog networks} {Efficient simplex-like methods for equilibria
  of nonsymmetric analog networks}.{\BBCQ}
\newblock
\APACjournalVolNumPages{Neural Computation}{4}{2}{167--190}.
\PrintBackRefs{\CurrentBib}

\bibitem [\protect \citeauthoryear {%
Miller%
\ \BBA {} Zucker%
}{%
Miller%
\ \BBA {} Zucker%
}{%
{\protect \APACyear {1999}}%
}]{%
MillerZucker1999}
\APACinsertmetastar {%
MillerZucker1999}%
\begin{APACrefauthors}%
Miller, D\BPBI A.%
\BCBT {}\ \BBA {} Zucker, S\BPBI W.%
\end{APACrefauthors}%
\unskip\
\newblock
\APACrefYearMonthDay{1999}{}{}.
\newblock
{\BBOQ}\APACrefatitle {Computing with self-excitatory cliques: a model and an
  application to hyperacuity-scale computation in visual cortex} {Computing
  with self-excitatory cliques: a model and an application to hyperacuity-scale
  computation in visual cortex}.{\BBCQ}
\newblock
\APACjournalVolNumPages{Neural Computation}{11}{1}{21--66}.
\PrintBackRefs{\CurrentBib}

\bibitem [\protect \citeauthoryear {%
Mostafa%
, M\"uller%
\BCBL {}\ \BBA {} Indiveri%
}{%
Mostafa%
\ \protect \BOthers {.}}{%
{\protect \APACyear {2013}}%
}]{%
Mostafa_etal13b}
\APACinsertmetastar {%
Mostafa_etal13b}%
\begin{APACrefauthors}%
Mostafa, H.%
, M\"uller, L\BPBI K.%
\BCBL {}\ \BBA {} Indiveri, G.%
\end{APACrefauthors}%
\unskip\
\newblock
\APACrefYearMonthDay{2013}{}{}.
\newblock
{\BBOQ}\APACrefatitle {Recurrent networks of coupled Winner-Take-All
  oscillators for solving constraint satisfaction problems} {Recurrent networks
  of coupled winner-take-all oscillators for solving constraint satisfaction
  problems}.{\BBCQ}
\newblock
\BIn{} \APACrefbtitle {Advances in Neural Information Processing Systems
  ({NIPS})} {Advances in neural information processing systems ({NIPS})}\
  (\BVOL~26, \BPGS\ 719--727).
\PrintBackRefs{\CurrentBib}

\bibitem [\protect \citeauthoryear {%
Nair%
\ \BBA {} Hinton%
}{%
Nair%
\ \BBA {} Hinton%
}{%
{\protect \APACyear {2010}}%
}]{%
nair2010rectified}
\APACinsertmetastar {%
nair2010rectified}%
\begin{APACrefauthors}%
Nair, V.%
\BCBT {}\ \BBA {} Hinton, G\BPBI E.%
\end{APACrefauthors}%
\unskip\
\newblock
\APACrefYearMonthDay{2010}{}{}.
\newblock
{\BBOQ}\APACrefatitle {Rectified linear units improve restricted boltzmann
  machines} {Rectified linear units improve restricted boltzmann
  machines}.{\BBCQ}
\newblock
\BIn{} \APACrefbtitle {Proceedings of the 27th international conference on
  machine learning (ICML-10)} {Proceedings of the 27th international conference
  on machine learning (icml-10)}\ (\BPGS\ 807--814).
\PrintBackRefs{\CurrentBib}

\bibitem [\protect \citeauthoryear {%
Neftci%
\ \protect \BOthers {.}}{%
Neftci%
\ \protect \BOthers {.}}{%
{\protect \APACyear {2013}}%
}]{%
Neftci2013}
\APACinsertmetastar {%
Neftci2013}%
\begin{APACrefauthors}%
Neftci, E.%
, Binas, J.%
, Rutishauser, U.%
, Chicca, E.%
, Indiveri, G.%
\BCBL {}\ \BBA {} Douglas, R\BPBI J.%
\end{APACrefauthors}%
\unskip\
\newblock
\APACrefYearMonthDay{2013}{}{}.
\newblock
{\BBOQ}\APACrefatitle {Synthesizing cognition in neuromorphic electronic
  systems} {Synthesizing cognition in neuromorphic electronic systems}.{\BBCQ}
\newblock
\APACjournalVolNumPages{Proceedings of the National Academy of
  Sciences}{110}{37}{E3468--E3476}.
\PrintBackRefs{\CurrentBib}

\bibitem [\protect \citeauthoryear {%
Neftci%
, Chicca%
, Indiveri%
\BCBL {}\ \BBA {} Douglas%
}{%
Neftci%
\ \protect \BOthers {.}}{%
{\protect \APACyear {2011}}%
}]{%
Neftci_etal11}
\APACinsertmetastar {%
Neftci_etal11}%
\begin{APACrefauthors}%
Neftci, E.%
, Chicca, E.%
, Indiveri, G.%
\BCBL {}\ \BBA {} Douglas, R\BPBI J.%
\end{APACrefauthors}%
\unskip\
\newblock
\APACrefYearMonthDay{2011}{}{}.
\newblock
{\BBOQ}\APACrefatitle {A systematic method for configuring {VLSI} networks of
  spiking neurons} {A systematic method for configuring {VLSI} networks of
  spiking neurons}.{\BBCQ}
\newblock
\APACjournalVolNumPages{Neural Computation}{23}{10}{2457--2497}.
\PrintBackRefs{\CurrentBib}

\bibitem [\protect \citeauthoryear {%
Niemeyer%
\ \BBA {} Slotine%
}{%
Niemeyer%
\ \BBA {} Slotine%
}{%
{\protect \APACyear {1997}}%
}]{%
NiemeyerSlotine1997}
\APACinsertmetastar {%
NiemeyerSlotine1997}%
\begin{APACrefauthors}%
Niemeyer, G.%
\BCBT {}\ \BBA {} Slotine, J.%
\end{APACrefauthors}%
\unskip\
\newblock
\APACrefYearMonthDay{1997}{}{}.
\newblock
{\BBOQ}\APACrefatitle {A simple strategy for opening an unknown door} {A simple
  strategy for opening an unknown door}.{\BBCQ}
\newblock
\BIn{} \APACrefbtitle {Robotics and Automation, 1997. Proceedings., 1997 IEEE
  International Conference on} {Robotics and automation, 1997. proceedings.,
  1997 ieee international conference on}\ (\BVOL~2, \BPGS\ 1448--1453).
\PrintBackRefs{\CurrentBib}

\bibitem [\protect \citeauthoryear {%
Pi%
\ \protect \BOthers {.}}{%
Pi%
\ \protect \BOthers {.}}{%
{\protect \APACyear {2013}}%
}]{%
Pi2013cortical}
\APACinsertmetastar {%
Pi2013cortical}%
\begin{APACrefauthors}%
Pi, H\BHBI J.%
, Hangya, B.%
, Kvitsiani, D.%
, Sanders, J\BPBI I.%
, Huang, Z\BPBI J.%
\BCBL {}\ \BBA {} Kepecs, A.%
\end{APACrefauthors}%
\unskip\
\newblock
\APACrefYearMonthDay{2013}{}{}.
\newblock
{\BBOQ}\APACrefatitle {Cortical interneurons that specialize in disinhibitory
  control} {Cortical interneurons that specialize in disinhibitory
  control}.{\BBCQ}
\newblock
\APACjournalVolNumPages{Nature}{503}{7477}{521--524}.
\PrintBackRefs{\CurrentBib}

\bibitem [\protect \citeauthoryear {%
\APACcitebtitle {Project Euler}}{%
\APACcitebtitle {Project Euler}}{%
{\protect \APACyear {2015}}%
}]{%
ProjectEuler}
\APACinsertmetastar {%
ProjectEuler}%
\APACrefbtitle {Project Euler.} {Project euler.}
\newblock
\APACrefYearMonthDay{2015}{}{}.
\newblock
\begin{APACrefURL} [{2015-08-27}]\url{https://projecteuler.net/problem=96}
  \end{APACrefURL}
\PrintBackRefs{\CurrentBib}

\bibitem [\protect \citeauthoryear {%
Qiao%
\ \protect \BOthers {.}}{%
Qiao%
\ \protect \BOthers {.}}{%
{\protect \APACyear {2015}}%
}]{%
qiao2015reconfigurable}
\APACinsertmetastar {%
qiao2015reconfigurable}%
\begin{APACrefauthors}%
Qiao, N.%
, Mostafa, H.%
, Corradi, F.%
, Osswald, M.%
, Stefanini, F.%
, Sumislawska, D.%
\BCBL {}\ \BBA {} Indiveri, G.%
\end{APACrefauthors}%
\unskip\
\newblock
\APACrefYearMonthDay{2015}{}{}.
\newblock
{\BBOQ}\APACrefatitle {A reconfigurable on-line learning spiking neuromorphic
  processor comprising 256 neurons and 128K synapses} {A reconfigurable on-line
  learning spiking neuromorphic processor comprising 256 neurons and 128k
  synapses}.{\BBCQ}
\newblock
\APACjournalVolNumPages{Frontiers in neuroscience}{9}{}{}.
\PrintBackRefs{\CurrentBib}

\bibitem [\protect \citeauthoryear {%
Robertson%
, Sanders%
, Seymour%
\BCBL {}\ \BBA {} Thomas%
}{%
Robertson%
\ \protect \BOthers {.}}{%
{\protect \APACyear {1996}}%
}]{%
Robertson1996}
\APACinsertmetastar {%
Robertson1996}%
\begin{APACrefauthors}%
Robertson, N.%
, Sanders, D\BPBI P.%
, Seymour, P.%
\BCBL {}\ \BBA {} Thomas, R.%
\end{APACrefauthors}%
\unskip\
\newblock
\APACrefYearMonthDay{1996}{}{}.
\newblock
{\BBOQ}\APACrefatitle {Efficiently four-coloring planar graphs} {Efficiently
  four-coloring planar graphs}.{\BBCQ}
\newblock
\BIn{} \APACrefbtitle {Proceedings of the twenty-eighth annual ACM symposium on
  Theory of computing} {Proceedings of the twenty-eighth annual acm symposium
  on theory of computing}\ (\BPGS\ 571--575).
\PrintBackRefs{\CurrentBib}

\bibitem [\protect \citeauthoryear {%
Rodger%
\ \BBA {} Finley%
}{%
Rodger%
\ \BBA {} Finley%
}{%
{\protect \APACyear {2006}}%
}]{%
Rodger06}
\APACinsertmetastar {%
Rodger06}%
\begin{APACrefauthors}%
Rodger, S.%
\BCBT {}\ \BBA {} Finley, T.%
\end{APACrefauthors}%
\unskip\
\newblock
\APACrefYear{2006}.
\newblock
\APACrefbtitle {JFLAP - An Interactive Formal Languages and Automata Package}
  {Jflap - an interactive formal languages and automata package}.
\newblock
\APACaddressPublisher{Sudbury, MA}{Jones and Bartlett}.
\newblock
\APACrefnote{ISBN 0763738344}
\PrintBackRefs{\CurrentBib}

\bibitem [\protect \citeauthoryear {%
Rosenfeld%
, Hummel%
\BCBL {}\ \BBA {} Zucker%
}{%
Rosenfeld%
\ \protect \BOthers {.}}{%
{\protect \APACyear {1976}}%
}]{%
RosenfeldZucker1976}
\APACinsertmetastar {%
RosenfeldZucker1976}%
\begin{APACrefauthors}%
Rosenfeld, A.%
, Hummel, R\BPBI A.%
\BCBL {}\ \BBA {} Zucker, S\BPBI W.%
\end{APACrefauthors}%
\unskip\
\newblock
\APACrefYearMonthDay{1976}{}{}.
\newblock
{\BBOQ}\APACrefatitle {Scene labeling by relaxation operations} {Scene labeling
  by relaxation operations}.{\BBCQ}
\newblock
\APACjournalVolNumPages{IEEE Transactions on Systems, Man, and
  Cybernetics}{6}{6}{420--433}.
\PrintBackRefs{\CurrentBib}

\bibitem [\protect \citeauthoryear {%
Rosenhouse%
\ \BBA {} Taalman%
}{%
Rosenhouse%
\ \BBA {} Taalman%
}{%
{\protect \APACyear {2011}}%
}]{%
Rosenhouse2011taking}
\APACinsertmetastar {%
Rosenhouse2011taking}%
\begin{APACrefauthors}%
Rosenhouse, J.%
\BCBT {}\ \BBA {} Taalman, L.%
\end{APACrefauthors}%
\unskip\
\newblock
\APACrefYear{2011}.
\newblock
\APACrefbtitle {Taking Sudoku Seriously: The Math Behind the Worlds Most
  Popular Pencil Puzzle} {Taking sudoku seriously: The math behind the worlds
  most popular pencil puzzle}.
\newblock
\APACaddressPublisher{}{Oxford University Press}.
\PrintBackRefs{\CurrentBib}

\bibitem [\protect \citeauthoryear {%
Rudy%
, Fishell%
, Lee%
\BCBL {}\ \BBA {} Hjerling-Leffler%
}{%
Rudy%
\ \protect \BOthers {.}}{%
{\protect \APACyear {2011}}%
}]{%
rudy_three_2011}
\APACinsertmetastar {%
rudy_three_2011}%
\begin{APACrefauthors}%
Rudy, B.%
, Fishell, G.%
, Lee, S.%
\BCBL {}\ \BBA {} Hjerling-Leffler, J.%
\end{APACrefauthors}%
\unskip\
\newblock
\APACrefYearMonthDay{2011}{}{}.
\newblock
{\BBOQ}\APACrefatitle {Three groups of interneurons account for nearly 100\% of
  neocortical {GABAergic} neurons} {Three groups of interneurons account for
  nearly 100\% of neocortical {GABAergic} neurons}.{\BBCQ}
\newblock
\APACjournalVolNumPages{Devel Neurobio}{71}{1}{45--61}.
\PrintBackRefs{\CurrentBib}

\bibitem [\protect \citeauthoryear {%
Russell%
\ \BBA {} Norvig%
}{%
Russell%
\ \BBA {} Norvig%
}{%
{\protect \APACyear {2010}}%
}]{%
russell_artificial_2010}
\APACinsertmetastar {%
russell_artificial_2010}%
\begin{APACrefauthors}%
Russell, S\BPBI J.%
\BCBT {}\ \BBA {} Norvig, P.%
\end{APACrefauthors}%
\unskip\
\newblock
\APACrefYear{2010}.
\newblock
\APACrefbtitle {Artificial intelligence: a modern approach (International
  Edition)} {Artificial intelligence: a modern approach (international
  edition)}\ (\PrintOrdinal{3rd}\ \BEd).
\newblock
\APACaddressPublisher{}{Prentice Hall}.
\PrintBackRefs{\CurrentBib}

\bibitem [\protect \citeauthoryear {%
Rutishauser%
\ \BBA {} Douglas%
}{%
Rutishauser%
\ \BBA {} Douglas%
}{%
{\protect \APACyear {2009}}%
}]{%
RutishauserDouglas2009}
\APACinsertmetastar {%
RutishauserDouglas2009}%
\begin{APACrefauthors}%
Rutishauser, U.%
\BCBT {}\ \BBA {} Douglas, R\BPBI J.%
\end{APACrefauthors}%
\unskip\
\newblock
\APACrefYearMonthDay{2009}{}{}.
\newblock
{\BBOQ}\APACrefatitle {State-Dependent Computation Using Coupled recurrent
  networks} {State-dependent computation using coupled recurrent
  networks}.{\BBCQ}
\newblock
\APACjournalVolNumPages{Neural Computation}{21}{2}{}.
\PrintBackRefs{\CurrentBib}

\bibitem [\protect \citeauthoryear {%
Rutishauser%
, Douglas%
\BCBL {}\ \BBA {} Slotine%
}{%
Rutishauser%
\ \protect \BOthers {.}}{%
{\protect \APACyear {2011}}%
}]{%
RutishauserDouglas2011}
\APACinsertmetastar {%
RutishauserDouglas2011}%
\begin{APACrefauthors}%
Rutishauser, U.%
, Douglas, R\BPBI J.%
\BCBL {}\ \BBA {} Slotine, J.%
\end{APACrefauthors}%
\unskip\
\newblock
\APACrefYearMonthDay{2011}{}{}.
\newblock
{\BBOQ}\APACrefatitle {Collective stability of networks of winner-take-all
  circuits} {Collective stability of networks of winner-take-all
  circuits}.{\BBCQ}
\newblock
\APACjournalVolNumPages{Neural computation}{23}{3}{}.
\PrintBackRefs{\CurrentBib}

\bibitem [\protect \citeauthoryear {%
Rutishauser%
, Slotine%
\BCBL {}\ \BBA {} Douglas%
}{%
Rutishauser%
\ \protect \BOthers {.}}{%
{\protect \APACyear {2012}}%
}]{%
rutishauser_competition_2012}
\APACinsertmetastar {%
rutishauser_competition_2012}%
\begin{APACrefauthors}%
Rutishauser, U.%
, Slotine, J.%
\BCBL {}\ \BBA {} Douglas, R\BPBI J.%
\end{APACrefauthors}%
\unskip\
\newblock
\APACrefYearMonthDay{2012}{}{}.
\newblock
{\BBOQ}\APACrefatitle {Competition through selective inhibitory synchrony}
  {Competition through selective inhibitory synchrony}.{\BBCQ}
\newblock
\APACjournalVolNumPages{Neural Computation}{24}{8}{2033--2052}.
\PrintBackRefs{\CurrentBib}

\bibitem [\protect \citeauthoryear {%
Rutishauser%
, Slotine%
\BCBL {}\ \BBA {} Douglas%
}{%
Rutishauser%
\ \protect \BOthers {.}}{%
{\protect \APACyear {2015}}%
}]{%
rutishauser_computation_2015}
\APACinsertmetastar {%
rutishauser_computation_2015}%
\begin{APACrefauthors}%
Rutishauser, U.%
, Slotine, J.%
\BCBL {}\ \BBA {} Douglas, R\BPBI J.%
\end{APACrefauthors}%
\unskip\
\newblock
\APACrefYearMonthDay{2015}{}{}.
\newblock
{\BBOQ}\APACrefatitle {Computation in Dynamically Bounded Asymmetric Systems}
  {Computation in dynamically bounded asymmetric systems}.{\BBCQ}
\newblock
\APACjournalVolNumPages{{PLoS} Comput Biol}{11}{1}{}.
\PrintBackRefs{\CurrentBib}

\bibitem [\protect \citeauthoryear {%
Siek%
, Lee%
\BCBL {}\ \BBA {} Lumsdaine%
}{%
Siek%
\ \protect \BOthers {.}}{%
{\protect \APACyear {2002}}%
}]{%
BGL2002}
\APACinsertmetastar {%
BGL2002}%
\begin{APACrefauthors}%
Siek, J\BPBI G.%
, Lee, L\BHBI Q.%
\BCBL {}\ \BBA {} Lumsdaine, A.%
\end{APACrefauthors}%
\unskip\
\newblock
\APACrefYear{2002}.
\newblock
\APACrefbtitle {The boost graph library: user guide and reference manual} {The
  boost graph library: user guide and reference manual}.
\newblock
\APACaddressPublisher{Boston, MA, USA}{Addison-Wesley Longman Publishing Co.,
  Inc.}
\PrintBackRefs{\CurrentBib}

\bibitem [\protect \citeauthoryear {%
Slotine%
}{%
Slotine%
}{%
{\protect \APACyear {2003}}%
}]{%
Slotine03}
\APACinsertmetastar {%
Slotine03}%
\begin{APACrefauthors}%
Slotine, J.%
\end{APACrefauthors}%
\unskip\
\newblock
\APACrefYearMonthDay{2003}{}{}.
\newblock
{\BBOQ}\APACrefatitle {Modular stability tools for distributed computation and
  control} {Modular stability tools for distributed computation and
  control}.{\BBCQ}
\newblock
\APACjournalVolNumPages{International Journal of Adaptive Control and Signal
  Processing}{17}{}{397-416}.
\PrintBackRefs{\CurrentBib}

\bibitem [\protect \citeauthoryear {%
Stuart%
\ \BBA {} Spruston%
}{%
Stuart%
\ \BBA {} Spruston%
}{%
{\protect \APACyear {2015}}%
}]{%
stuart_dendritic_2015}
\APACinsertmetastar {%
stuart_dendritic_2015}%
\begin{APACrefauthors}%
Stuart, G\BPBI J.%
\BCBT {}\ \BBA {} Spruston, N.%
\end{APACrefauthors}%
\unskip\
\newblock
\APACrefYearMonthDay{2015}{}{}.
\newblock
{\BBOQ}\APACrefatitle {Dendritic integration: 60 years of progress} {Dendritic
  integration: 60 years of progress}.{\BBCQ}
\newblock
\APACjournalVolNumPages{Nature Neuroscience}{18}{12}{1713--1721}.
\PrintBackRefs{\CurrentBib}

\bibitem [\protect \citeauthoryear {%
Tran-Van-Minh%
\ \protect \BOthers {.}}{%
Tran-Van-Minh%
\ \protect \BOthers {.}}{%
{\protect \APACyear {2015}}%
}]{%
tran-van-minh_contribution_2015}
\APACinsertmetastar {%
tran-van-minh_contribution_2015}%
\begin{APACrefauthors}%
Tran-Van-Minh, A.%
, Cazé, R\BPBI D.%
, Abrahamsson, T.%
, Cathala, L.%
, Gutkin, B\BPBI S.%
\BCBL {}\ \BBA {} {DiGregorio}, D\BPBI A.%
\end{APACrefauthors}%
\unskip\
\newblock
\APACrefYearMonthDay{2015}{}{}.
\newblock
{\BBOQ}\APACrefatitle {Contribution of sublinear and supralinear dendritic
  integration to neuronal computations} {Contribution of sublinear and
  supralinear dendritic integration to neuronal computations}.{\BBCQ}
\newblock
\APACjournalVolNumPages{Front Cell Neurosci}{9}{}{}.
\PrintBackRefs{\CurrentBib}

\bibitem [\protect \citeauthoryear {%
C\BPBI J.~Wang%
\ \BBA {} Tsang%
}{%
C\BPBI J.~Wang%
\ \BBA {} Tsang%
}{%
{\protect \APACyear {1991}}%
}]{%
wang1991solving}
\APACinsertmetastar {%
wang1991solving}%
\begin{APACrefauthors}%
Wang, C\BPBI J.%
\BCBT {}\ \BBA {} Tsang, E\BPBI P.%
\end{APACrefauthors}%
\unskip\
\newblock
\APACrefYearMonthDay{1991}{}{}.
\newblock
{\BBOQ}\APACrefatitle {Solving constraint satisfaction problems using neural
  networks} {Solving constraint satisfaction problems using neural
  networks}.{\BBCQ}
\newblock
\BIn{} \APACrefbtitle {Artificial Neural Networks, 1991., Second International
  Conference on} {Artificial neural networks, 1991., second international
  conference on}\ (\BPGS\ 295--299).
\PrintBackRefs{\CurrentBib}

\bibitem [\protect \citeauthoryear {%
W.~Wang%
\ \BBA {} Slotine%
}{%
W.~Wang%
\ \BBA {} Slotine%
}{%
{\protect \APACyear {2006}}%
}]{%
wang2006fast}
\APACinsertmetastar {%
wang2006fast}%
\begin{APACrefauthors}%
Wang, W.%
\BCBT {}\ \BBA {} Slotine, J.%
\end{APACrefauthors}%
\unskip\
\newblock
\APACrefYearMonthDay{2006}{}{}.
\newblock
{\BBOQ}\APACrefatitle {Fast computation with neural oscillators} {Fast
  computation with neural oscillators}.{\BBCQ}
\newblock
\APACjournalVolNumPages{Neurocomputing}{69}{16}{2320--2326}.
\PrintBackRefs{\CurrentBib}

\bibitem [\protect \citeauthoryear {%
Wersing%
, Steil%
\BCBL {}\ \BBA {} Ritter%
}{%
Wersing%
\ \protect \BOthers {.}}{%
{\protect \APACyear {2001}}%
}]{%
wersing2001competitive}
\APACinsertmetastar {%
wersing2001competitive}%
\begin{APACrefauthors}%
Wersing, H.%
, Steil, J\BPBI J.%
\BCBL {}\ \BBA {} Ritter, H.%
\end{APACrefauthors}%
\unskip\
\newblock
\APACrefYearMonthDay{2001}{}{}.
\newblock
{\BBOQ}\APACrefatitle {A competitive-layer model for feature binding and
  sensory segmentation} {A competitive-layer model for feature binding and
  sensory segmentation}.{\BBCQ}
\newblock
\APACjournalVolNumPages{Neural Computation}{13}{2}{357--387}.
\PrintBackRefs{\CurrentBib}

\bibitem [\protect \citeauthoryear {%
Wu%
\ \BBA {} Hao%
}{%
Wu%
\ \BBA {} Hao%
}{%
{\protect \APACyear {2015}}%
}]{%
wu_review_2015}
\APACinsertmetastar {%
wu_review_2015}%
\begin{APACrefauthors}%
Wu, Q.%
\BCBT {}\ \BBA {} Hao, J\BHBI K.%
\end{APACrefauthors}%
\unskip\
\newblock
\APACrefYearMonthDay{2015}{}{}.
\newblock
{\BBOQ}\APACrefatitle {A review on algorithms for maximum clique problems} {A
  review on algorithms for maximum clique problems}.{\BBCQ}
\newblock
\APACjournalVolNumPages{European Journal of Operational
  Research}{242}{3}{693--709}.
\PrintBackRefs{\CurrentBib}

\bibitem [\protect \citeauthoryear {%
Yuille%
\ \BBA {} Geiger%
}{%
Yuille%
\ \BBA {} Geiger%
}{%
{\protect \APACyear {2003}}%
}]{%
YuilleGeiger03}
\APACinsertmetastar {%
YuilleGeiger03}%
\begin{APACrefauthors}%
Yuille, A.%
\BCBT {}\ \BBA {} Geiger, D.%
\end{APACrefauthors}%
\unskip\
\newblock
\APACrefYearMonthDay{2003}{}{}.
\newblock
{\BBOQ}\APACrefatitle {Winner-Take-All Networks} {Winner-take-all
  networks}.{\BBCQ}
\newblock
\BIn{} M.~Arbib\ (\BED), \APACrefbtitle {The Handbook of Brain Theory and
  Neural Networks} {The handbook of brain theory and neural networks}\
  (\BPG~1228-1231).
\newblock
\APACaddressPublisher{}{MIT Press}.
\PrintBackRefs{\CurrentBib}

\bibitem [\protect \citeauthoryear {%
Zhang%
, Li%
, Rasch%
\BCBL {}\ \BBA {} Wu%
}{%
Zhang%
\ \protect \BOthers {.}}{%
{\protect \APACyear {2013}}%
}]{%
zhang_nonlinear_2013}
\APACinsertmetastar {%
zhang_nonlinear_2013}%
\begin{APACrefauthors}%
Zhang, D.%
, Li, Y.%
, Rasch, M\BPBI J.%
\BCBL {}\ \BBA {} Wu, S.%
\end{APACrefauthors}%
\unskip\
\newblock
\APACrefYearMonthDay{2013}{}{}.
\newblock
{\BBOQ}\APACrefatitle {Nonlinear multiplicative dendritic integration in neuron
  and network models} {Nonlinear multiplicative dendritic integration in neuron
  and network models}.{\BBCQ}
\newblock
\APACjournalVolNumPages{Front Comput Neurosci}{7}{}{}.
\PrintBackRefs{\CurrentBib}

\end{thebibliography}

\end{document}